\documentclass{article}
\usepackage{graphicx} 
\usepackage{amsmath}
\usepackage{amsthm}
\usepackage{amssymb}
\usepackage{mathtools}
\usepackage{xcolor}
\usepackage[round]{natbib}
\usepackage{enumerate}
\usepackage{bbm}
\usepackage{physics}
\usepackage{apptools}
\usepackage{float}
\usepackage{threeparttable}
\usepackage{rotating} 

\usepackage[a4paper, total={6in, 8in}]{geometry}

\usepackage{authblk}

\usepackage{orcidlink}

\AtAppendix{\counterwithin{theorem}{section}}

\newtheorem{theorem}{Theorem}
\newtheorem{lemma}[theorem]{Lemma}
\newtheorem*{remark}{Remark}
\newtheorem{corollary}[theorem]{Corollary}

\newtheorem{assumption}[theorem]{Assumption}

\theoremstyle{definition}

\providecommand{\keywords}[1]
{
  \small	
  \textbf{\textit{Keywords---}} #1
}

\newcommand\extrafootertext[1]{%
    \bgroup
    \renewcommand\thefootnote{\fnsymbol{footnote}}%
    \renewcommand\thempfootnote{\fnsymbol{mpfootnote}}%
    \footnotetext[0]{#1}%
    \egroup
}

\title{Inference in pseudo-observation-based regression using (biased) covariance estimation and naive bootstrapping}

\author{
Simon Mack $^{1,\ast}$\orcidlink{0009-0006-0081-935X},
Morten Overgaard $^{2}$\orcidlink{0000-0001-8052-0337} and
Dennis Dobler $^{1}$\orcidlink{0000-0002-9040-0854}}

\affil{
$^{1}$ Institute of Statistics, RWTH Aachen University, Aachen, Germany \\
$^{2}$ Department of Public Health - Department of Biostatistics, Aarhus University, Aarhus, Denmark}

\date{\today}

\begin{document}

\maketitle

\begin{abstract}
The pseudo-observation method is regularly applied to time-to-event data. However, to date such analyses have relied on not formally verified statements or ad-hoc methods regarding covariance estimation. This paper strives to close this gap in the literature. To begin with, we demonstrate that the usual Huber-White estimator is not consistent for the limiting covariance of parameter estimates in pseudo-observation regression approaches.
By confirming that a plug-in estimator can be used instead,
we obtain asymptotically exact and consistent tests for general linear hypotheses in the parameters of the model. Additionally, we confirm that naive bootstrapping
can not be used for covariance estimation in the pseudo-observation model either.
However, it can be used for hypothesis testing by applying a suitable
studentization. Simulations illustrate the good performance of our proposed methods in many scenarios. Finally, we obtain a general uniform law of large numbers for
U- and V-statistics, as such statistics are central in the mathematical analysis of the inference procedures developed in this work.
\end{abstract}


\extrafootertext{$\ast$ Corresponding author. Email adress: \texttt{simon.mack@rwth-aachen.de}}
\section{Introduction}
 The pseudo-observation regression approach, 
 introduced by \citet{andersen2003generalised}, provides a flexible alternative to the omnipresent proportional hazards model \citep{cox1972regression} when modeling time-to-event outcomes. In this approach, estimands representable as expectations are fitted to regression models using covariates of interest. 
 Exemplary estimands that fit this framework are the restricted mean time lost (in competing risks models) or the survival function at a fixed time-point (in simply survival models); cf.\ \citet{overgaard2017}.
 For fully observable data,
 the typical approach to estimate the parameters of a regression model is to define the estimator as the root of an estimating function. 
 This is not possible if observability of the response variable is restricted, for example due to censoring. In the pseudo-observation regression approach, the only partially observable responses are replaced by leave-one-out jackknife replicates based on an estimator for the unconditional mean of the response.

To investigate conditions under which the proposed method of \citet{andersen2003generalised} is valid, \citet{graw2009pseudo} studied the pseudo-observations in the competing risks setting in terms of so called von-Mises expansions. 
 However, it turned out that the remainder terms involved in this expansion cannot be ignored;
 while all other conclusions of \citet{graw2009pseudo} remain valid, the asymptotic variance of the estimator contains second-order terms and is in general overestimated by the proposed Huber-White-type sandwich estimator of \citet{andersen2003generalised}; see \citet{jacobsen2016note} for details. 
 \citet{overgaard2017} extended these results and obtained detailed conditions for the validity of the pseudo-observation approach, as well as the limiting variance of the estimators in the general case. They also conjectured that the usual sandwich estimator fails to correctly estimate the variance in general; instead, they suggested a plug-in estimator. Simulations suggest that this estimator performs well in many scenarios \citep{overgaard2018estimating} but a mathematical proof of its consistency is not yet available. Another approach to variance estimation is the bootstrap; it offers another alternative to the Huber-White estimator.
Although this was suggested more than twenty years ago \citep{andersen2003generalised}, the mathematical validity of such a procedure has still not been investigated in the present context either.

 One benefit of consistent (co-)variance estimators is the possibility to construct 
 tests for general linear hypotheses in the parameters of the model. 
Such hypotheses frequently arise in regression models with categorical covariates, for example to test main and interaction effects in factorial designs. 
The duality between testing and confidence regions would also allow for multivariate confidence ellipsoids for the parameter vector.

In this article we fill these gaps by proving the consistency of the plug-in variance estimator of \citet{overgaard2017} and by proposing a quadratic form statistic of Wald-type to test general linear hypotheses. This constitutes an extension of tests and confidence intervals for univariate regression parameters which are commonly applied in the literature; however, their asymptotic properties were not formally verified yet. We also take a closer look at a naive approach for bootstrapping pseudo-observations \citep{andersen2003generalised}; it turns out to be invalid for variance estimation.
However, we investigate its potential validity for hypothesis testing after applying a 
suitable standardization.

The remainder of the article is organized as follows. In Section, \ref{sec:model and notation} we introduce the inferential problems of interest and give a detailed introduction to the pseudo-observation method. Section \ref{sec:covariance estimation and hypothesis tests} contains our main  results on \mbox{(co-)}variance estimation and tests for general linear hypotheses. A naive bootstrap approach is investigated in Section \ref{sec:bootstrap}. In Section \ref{sec:simulation and data example}, the type-I error and power of the obtained hypothesis tests is assessed in a simulation study and an illustrative data analysis is performed. We conclude with a discussion in Section \ref{sec:discussion}.
In Appendix \ref{appendix:A uniform LLN} we present general uniform laws of large numbers for U- and V-statistics, which is central to the asymptotic analysis of the regression parameter and covariance estimators as well as the resampling approaches considered in this paper; these analyses as well as all proofs can be found in Appendix \ref{appendix:B proofs}. To the best of our knowledge, such a uniform law of large numbers is not yet available in the literature. Appendix \ref{appendix:C add simulation} contains further simulation results.

\section{Model, notation, and previous results}\label{sec:model and notation}
To formalize the statistical model and  approach, we closely follow the notation of \citet{overgaard2017}.
Let   $(V, X, Z)$ be a triplet of $(\mathbb{R}\times \mathcal{X}\times\mathcal{Z})$-valued random variables on a probability space $(\Omega, \mathcal{F}, P)$; in typical applications, $\mathcal X$ and $\mathcal Z$ are Euclidean spaces.   
The response variable $V$ is usually not fully observable, $Z$ represents observable covariates assuming the role of explanatory variables, and $X$ are observable additional variables enabling the estimation of $E(V)$.

We assume our estimand of interest is  of the form
\[E(V|Z)=\mu(\beta_0, Z)\]
for a mean function $\mu$ which usually is a response function from a generalized linear model and the true parameter vector $\beta_0 \in \Theta \subset \mathbb{R}^q$. In most cases, abusing notation $\mu(\beta_0, Z)= \mu(\beta_0^TZ)$. Throughout the paper, we assume the parameter space $\Theta$ to be open. Our goal is the estimation of $\beta_0$ and the construction of valid tests for general linear hypothesis of the form 
\begin{equation}\label{generalhypothesis}
    H_0:C\beta_0=b \text{ vs. } H_1:C\beta_0\ne b
\end{equation}
for known matrices $C\in \mathbb{R}^{p\times q}$ and vectors $b \in \mathbb{R}^p$. Consider independent and identically distributed (i.i.d.)\  tuples $(V_1, X_1, Z_1), \dots, (V_n, X_n, Z_n)$ which are copies of $(V,X,Z)$.

Let $D$ be some Banach space with norm $||\cdot||_D$.  Assume there exists a consistent estimator of the unconditional expected value $\theta= E(V)$ of the form $\hat{\theta}_n=\phi(F_n)$ based on a sample average $F_n=\frac{1}{n}\sum_{k=1}^n \delta_{X_k} \in D$
where $\delta_{(.)}: \mathcal X \to D$ is a map and a functional $\phi:D_0 \to \mathbb{R}$ defined on some suitable subspace $D_0\subset D$. When modeling time-to-event outcomes, $X_i$ typically represents the event times and censoring indicators $\delta_{X_i}$ the vector of at-risk and counting processes; cf.\ Example~2.1 in \citet{overgaard2017} for more details in this context.
This setup allows for the definition of jackknife pseudo-observations, which are given by
\begin{equation*}
    \hat{\theta}_{n,k}= n\hat{\theta}_n-(n-1)\hat{\theta}_n^{(k)}=n\phi(F_n)-(n-1)\phi(F_n^{(k)})
\end{equation*}
with $F_n^{(k)}=\frac{1}{n-1}\sum_{i\ne k}\delta_{X_i}$ denoting the leave-$1$-out average, i.e., without the $k$th observation. One method to estimate $\beta_0$ for fully observable data is to define the estimator as a solution to the estimating equation
\begin{equation}\label{fullequation}
    \sum_{k=1}^n A(\beta, Z_k)(V_k-\mu(\beta, Z_k))=0.
\end{equation}
The function $A(\beta, Z) \in \mathbb{R}^q$ is usually proportional to $\frac{\partial}{\partial\beta}\mu(\beta, Z)$. \citet{andersen2003generalised} proposed to use the pseudo-observations $(\hat{\theta}_{n,1}, \dots, \hat{\theta}_{n,n})$ to solve the problem of the (partial) unobservability of the responses $(V_1, \dots, V_n)$. This leads to the estimating equation
\begin{equation}\label{pseudo_equation}
    \sum_{k=1}^n A(\beta, Z_k)(\hat{\theta}_{n,k} -\mu(\beta, Z_k))=0.
\end{equation}
\citet{overgaard2017} established that  $\hat \beta$ is consistent for $\beta_0$ and asymptotically normal but it exhibits an altered asymptotic covariance structure in comparison to estimators obtained from \eqref{fullequation} for completely observable data.

Most of the asymptotic analysis of pseudo-observations relies on the following regularity assumptions on the estimating functional.
\begin{assumption}\label{assumption:functional}
    There exists an element $F \in D$ such that
    \begin{enumerate}[(a)]
        \item $||F-F_n||_D=o_p(n^{-\frac{1}{4}})$ and $||\delta_x||_D \leq c$ for some $0<c<\infty$ and for all $x \in \mathcal{X}$,
        \item $\phi$ is twice Fréchet-differentiable in a neighborhood of $F$ and the second-order derivative is Lip\-schitz continuous in a neighborhood of F.
    \end{enumerate}
\end{assumption}
This assumption is fulfilled for many functionals relevant to survival analysis. To this end, one can choose $D$ to be the space of functions on an interval $J \subset \mathbb R$ with finite $p$-variation norm cf.\ \citet{overgaard2017, dudley1999differentiability, dudley2011concrete} for details.

Based on this assumption, \citet{overgaard2017} obtained a decomposition of the pseudo-observations

    \begin{equation*}
        \hat{\theta}_{n,k}= \hat{\theta}^{\ast}_{n,k} + R_{n,k}
    \end{equation*}
     with 
     \begin{equation}\label{essentialpart}
         \hat{\theta}^{\ast}_{n,k}=\phi(F)+\phi_F'(\delta_{X_k}-F)+\phi_F''(\delta_{X_k}-F, F_n^{(k)}-F)
     \end{equation} 
     and $\sqrt{n}\max_k |R_{n,k}|=o_p(1)$.
Denoting $\Dot{\phi}(x)=\phi_F'(\delta_x-F)$ and $\Ddot{\phi}(x_1,x_2)=\phi_F''(\delta_{x_1}-F, \delta_{x_2}-F)$ the first- and second-order influence functions, one can rewrite the so-called \emph{essential part} $\hat{\theta}^{\ast}_{n,k}$ of $\hat \theta_{n,k}$ as
\begin{equation*}
     \hat{\theta}^{\ast}_{n,k}=\phi(F) + \Dot{\phi}(X_k)+ \frac{1}{n-1}\sum_{i\ne k}\Ddot{\phi}(X_k, X_i).
\end{equation*}
Based on this representation, \citet{overgaard2017} also established the identity $E(\hat{\theta}^{\ast}_{n,k} | Z_k)= \phi(F)+ E(\Dot{\phi}(X_k)|Z_k)$. Therefore, a sufficient condition for the estimating equation \eqref{pseudo_equation} to be asymptotically unbiased, is $E(\hat{\theta}^{\ast}_{n,k} | Z_k)=\mu(\beta_0, Z_k)$ which is implied by the following assumption if the mean function is correctly specified.
\begin{assumption}\label{assumption:centering}
    \begin{equation*}\label{condEcondition}
    E(\Dot{\phi}(X)|Z) = E(V|Z)-\phi(F).
\end{equation*}
\end{assumption}

For example, in the context of survival analysis \citet{overgaard2019pseudo} demonstrated that this assumption is fulfilled in the case of conditionally independent censoring if certain estimators that have an inverse probability of censoring weighting representation are considered. To elaborate, the nonparametric estimator of $\theta$ is of the form
\begin{equation*}
    \hat{\theta}_{n}=\frac{1}{n}\sum_{k=1}^n V_k\frac{\mathbbm{1}(C_k\geq \min(T_k,t))}{\hat{G}_n(\min(T_k,t)|Z_k)}.
\end{equation*}
Here, $(T_1,C_1,Z_1), \dots, (T_n,C_n,Z_n)$ are i.i.d., $T$ non-negative event times, $C$ non-negative censoring times, $t$ a fixed time point of interest, $Z$ covariates influencing the censoring distribution, and $\hat{G}_n$ an estimator for the survival function $G$ of the censoring random variables $C$. The assumption of conditional independence translates to the independence of $T$ and $C$ conditional on the covariates $Z.$ A common choice for $\hat{G}_n$ is the Kaplan-Meier estimator \citep{kaplan1958nonparametric} stratified by covariate levels if the assumption is made that the censoring distribution only depends on categorical covariates. No stratification corresponds to the assumption of completely independent censoring. Other common and suitable choices for the censoring distribution are the proportional hazards model \citep{cox1972regression} or the additive hazard regression model  \citep{aalen1989linear} as long as the support of the covariates is suitably restricted; cf.\ \citet{overgaard2019pseudo} for details.

For the remainder of this paper, we make the Assumptions \ref{assumption:functional} and \ref{assumption:centering}.
Based on these observations, \citet{overgaard2017} obtained the asymptotic normality of
\begin{equation}
\label{clt_ee}
  \frac{1}{\sqrt{n}}\sum_{k=1}^n A(\beta_0, Z_k)(\hat{\theta}_{n,k} -\mu(\beta_0, Z_k))  
\end{equation}
by recognizing that the asymptotically equivalent expression
\begin{equation*}
    \frac{1}{\sqrt{n}}\sum_{k=1}^n A(\beta_0, Z_k)(\hat{\theta}^{\ast}_{n,k} - E(\hat{\theta}^{\ast}_{n,k} | Z_k))
\end{equation*}
is an appropriately rescaled second-order U-statistic.
Based on this, also the asymptotic covariance matrix could be derived.

Another consequence of the central limit theorem for \eqref{clt_ee} is another central limit theorem for the solution $\hat \beta_n$ of the estimating equation~\eqref{pseudo_equation}:
$\sqrt{n}(\Hat{\beta}_n-\beta_0)$ is asymptotically normal with mean 0 and covariance matrix $M^{-1}\Sigma (M^{-1})^T$ where
\begin{equation*}
    M=-E\left(A(\beta_0,Z)\frac{\partial}{\partial\beta}\mu(\beta,Z)_{|\beta=\beta_0} \right),
\end{equation*}
\begin{equation*}\label{meat part limiting variance}
        \begin{split}
            \Sigma=\text{Cov}\Big(A(\beta_0,Z_1)&\big(\phi(F)+\dot\phi(X_1)-\mu(\beta_0,Z_1)\big)\\&+ E(A(\beta_0,Z_2)\ddot\phi(X_2,X_1)|X_1)\Big).
        \end{split}
\end{equation*}

\section{Covariance estimation and hypothesis tests}\label{sec:covariance estimation and hypothesis tests}
In order to test hypotheses about the parameter vector $\beta_0$, it is necessary to choose suitable test statistics. For the testing problem \eqref{generalhypothesis} we propose the Wald-type statistic
\begin{equation*}
    T_n=n(C\hat{\beta}_n-b)^T(C\Hat{M}_n^{-1}\Hat{\Sigma}_n(\Hat{M}_n^{-1})^TC^T)^{+}(C\hat{\beta}_n-b).
\end{equation*}
Here $H^+$ denotes the Moore-Penrose inverse of a matrix $H$ and $\Hat{M}_n$ and $\Hat{\Sigma}_n$ are estimators of $M$ and $\Sigma$, respectively.
Denote by $\stackrel d \to $ and $\stackrel p \to$ convergence in distribution and in probability, respectively, as $n \to \infty$.

The matrix $M$ is readily estimable with the simple plug-in estimator
        \begin{equation}\label{M Hat}
        \Hat{M}_n\coloneqq - \frac{1}{n} \sum_{k=1}^n A(\Hat{\beta}_n, Z_k)\frac{\partial}{\partial\beta}\mu(\beta, Z_k)_{|\beta=\Hat{\beta}_n},
    \end{equation}
as demonstrated in the subsequent lemma.
\begin{lemma}\label{M estimator}
    Assume that the following regularity conditions are fulfilled:
    \begin{enumerate}
        \item The functions $A(\beta, z)$ and $\frac{\partial}{\partial\beta}\mu(\beta, z)$ are continuously differentiable w.r.t.\ $\beta$ for all $z \in \mathcal{Z}$,
        \item $A(\beta, Z)\frac{\partial}{\partial\beta}\mu(\beta, Z)$ is dominated integrable in a neighborhood of $\beta_0,$
        \item $\Hat{\beta}_n\xrightarrow{p}\beta_0$ as $n\to\infty$.
        \end{enumerate}
    Then, as $n \to \infty$, we have $\Hat{M}_n \xrightarrow{p}M.$
\end{lemma}

On the other hand, the estimation of $\Sigma$ poses more difficulties. \citet{andersen2003generalised} suggested to use the Huber-White-type estimator
\begin{equation*}
    \Hat{\Sigma}^{\text{HW}}_n=\frac{1}{n}\sum_{k=1}^n A(\Hat{\beta}_n, Z_k)A(\Hat{\beta}_n, Z_k)^T(\hat{\theta}_{n,k} -\mu(\Hat{\beta}_n, Z_k))^2.
\end{equation*}
 However, \citet{overgaard2017} conjectured  that this estimator is not consistent for $\Sigma$.
Indeed, this is confirmed in the following theorem.
\begin{theorem}\label{HW convergence}
Assume that the following regularity conditions are fulfilled:
        \begin{enumerate}
        \item The functions $A(\beta, z)$ and $\mu(\beta, z)$ are continuous in $\beta$ for all $z \in \mathcal{Z}$,
        \item  $A(\beta, Z)$ and $A(\beta, Z)\mu(\beta, Z)$ are dominated square integrable in a neighborhood of $\beta_0,$
        \item $\Hat{\beta}_n\xrightarrow{p}\beta_0$ as $n\to\infty$.
    \end{enumerate}
    Then, as $n\to\infty$, we have
    \[\Hat{\Sigma}^{\text{HW}}_n \xrightarrow{p}\text{Cov}(A(\beta_0, Z_1)(\phi(F)+\dot \phi(X_1)-\mu(\beta, Z_1)))\eqcolon \Tilde{\Sigma}.\]
\end{theorem}

In general, $\Sigma$ and $\Tilde{\Sigma}$ are not identical because $\tilde \Sigma$ ignores the second derivative $\ddot \phi$ induced by the essential part of the pseudo-observations.
Furthermore, the diagonal elements of $\tilde \Sigma$ often exceed those of $\Sigma$  \citep{jacobsen2016note, overgaard2018estimating, overgaard2025comparison}. Therefore, the Huber-White estimator is biased. As the true variance is overestimated, hypothesis tests based on this estimator may lose a considerable amount of power. This is a particular problem in fields such as medicine where obtaining large samples can be challenging or expensive.

\citet{overgaard2017} proposed an alternative estimator for $\Sigma$:
\begin{equation}\label{corrected Var estimator}
\Hat{\Sigma}^{PV}_n := \frac{1}{n}\sum_{k=1}^n\left(A(\Hat{\beta}_n, Z_k)\left((\phi(F_n)+\phi_{F_n}'(\delta_{X_k}-F_n) -\mu(\Hat{\beta}_n,Z_k)\right) +\Hat{h}_1(X_k)\right)^{\otimes 2}
\end{equation}
where $\Hat{h}_1(x)=\frac{1}{n}\sum_{j=1}^nA(\Hat{\beta}_n,Z_j)\phi_{F_n}''(\delta_x-F_n,\delta_{X_j}-F_n)$ and $a^{\otimes 2}:=aa^T$, based on a Euclidean vector $a$.
The following theorem demonstrates that this estimator is consistent under the same conditions as in Theorem \ref{HW convergence}.
\begin{theorem}\label{Consistency corrected variance est}
Make the assumptions of Theorem \ref{HW convergence}.
Then, as $n \to \infty$,  we have $\Hat{\Sigma}^{PV}\xrightarrow{p}\Sigma.$
\end{theorem}

From now on we only consider $T_n$ based on the estimators $\Hat{M}_n$ from \eqref{M Hat} and $\Hat{\Sigma}^{PV}$ from \eqref{corrected Var estimator} if not specified otherwise. We can now formulate our main convergence result for the Wald-type statistic under the combined previous assumptions.

\begin{theorem}\label{asymptotic Wald test}
    Assume that the following regularity conditions are met:
    \begin{enumerate}
        \item $\mu(\beta, z)$ and $A(\beta,z)$ are continuously differentiable w.r.t.\ $\beta$ for all $z\in \mathcal{Z}$,
        \item  $A(\beta, Z)$ and $A(\beta, Z)\mu(\beta, Z)$ are dominated square integrable in a neighborhood of $\beta_0,$
        \item $\frac{\partial}{\partial\beta}(A(\beta,Z))\mu(\beta,Z), A(\beta,Z)\frac{\partial}{\partial\beta}\mu(\beta,Z)$ and $\frac{\partial}{\partial\beta}A(\beta,Z)$ are dominated integrable in a neighborhood of $\beta_0$,
        \item the matrix
        \begin{equation*}
            M=-E\left(A(\beta_0,Z)\frac{\partial}{\partial\beta}\mu(\beta,Z)_{|\beta=\beta_0} \right)
        \end{equation*}
        is invertible.
    \end{enumerate}
    Additionally assume that the linear system of equations $C\beta_0=b$ has at least one solution. Then, as $n\to\infty$, we have
    \begin{enumerate}
        \item[(a)] $T_n\xrightarrow{d}\chi^2_{\text{rank}(C)}$ under $H_0.$
        \item[(b)] $T_n\xrightarrow{p}\infty$  under $H_1.$
    \end{enumerate}
\end{theorem}
\begin{remark}
    For many mean functions (for example identity, inverse logit and inverse cloglog), the dominated integrability conditions are fulfilled if the covariates have finite second moments.
\end{remark}
Thus, we obtain an asymptotically valid level-$\alpha$-test
\begin{equation*}
    \varphi_{n,\alpha}=\mathbbm{1}\{T_n> q_{\text{rank}(C),1-\alpha}\}
\end{equation*}
where $q_{\text{rank}(C),1-\alpha}$ denotes the $(1-\alpha)$-quantile of the $\chi^2_{\text{rank}(C)}$-distribution, $\alpha\in (0,1).$ In mathematical terms, $\lim_{n\to \infty}E_{H_0}(\varphi_{n,\alpha})=\alpha$  and $\lim_{n\to \infty}E_{H_1}(\varphi_{n,\alpha})=1.$

These results demonstrate that hypothesis tests based on the Wald-type statistic can be applied in practice as long as the corrected variance estimator $\Hat{\Sigma}_n^{PV}$ is available for the estimand of interest. From a theoretical perspective this is of no concern but especially the explicit computation of the second-order influence function may be cumbersome in applications.
For example, we refer to Sections~3 and~4 in \citet{overgaard2018estimating} for the elaborate and explicit formulas of the theoretical covariances and the resulting plug-in estimators related to the Aalen-Johansen estimator in competing risks models. Such derivations might bear the risk of mathematical errors and flaws in software implementations when the involved functionals are very complex.

\section{Naive bootstrap approach}\label{sec:bootstrap}
Due to the complex limiting distribution of the parameter estimates, an appropriate resampling approach offers itself as an attractive alternative for variance estimation. \citet{andersen2003generalised} suggested to obtain bootstrap samples $(\hat{\theta}_{n,1}^B,Z_1^B),\dots, (\hat{\theta}_{n,n}^B, Z_n^B)$ randomly by sampling with replacement from the tuples $(\hat{\theta}_{n,1},Z_1),\dots, (\hat{\theta}_{n,n}, Z_n)$. 
Denoting by $\hat{\beta}_n^B$ the solution of the resulting estimating equation,
\begin{equation*}\label{pseudo_equation_bs}
    \sum_{k=1}^n A(\beta, Z_k^B)(\hat{\theta}_{n,k}^B -\mu(\beta, Z_k^B))=0,
\end{equation*}
the bootstrap can be used for variance estimation if
\begin{equation}\label{bootstrapvariance}
    E\left(n(\hat{\beta}_n^B-\hat{\beta}_n)^{\otimes2}|(\hat{\theta}_{n,1},Z_1),\dots, (\hat{\theta}_{n,n}, Z_n)\right)
\end{equation}
is consistent for $M^{-1}\Sigma (M^{-1})^T$ as $n\to\infty.$
We investigate the convergence properties of the bootstrap estimator in the subsequent theorem.

\begin{theorem}\label{weakconvbootstrapestimator}
Let $(M_{n1}, \dots, M_{nn}) \sim Mult(n, \frac{1}{n}, \dots, \frac{1}{n})$ be a multinomial random vector, independent of the data. Rewrite the bootstrap estimating equation as
    \begin{equation*}\label{B_estimatingequation}
        U_n^B(\beta)=\sum_{k=1}^n M_{nk} A(\beta, Z_k)(\hat{\theta}_{n,k} -\mu(\beta, Z_k)).
    \end{equation*}
    Assume that the following regularity conditions are met:
    \begin{enumerate}
        \item $\mu(\beta, z)$ and $A(\beta,z)$ are continuously differentiable w.r.t. $\beta$ for all $z\in \mathcal{Z}$,
        \item $A(\beta_0, Z)$ has finite second moment,
        \item $\frac{\partial}{\partial\beta}(A(\beta,Z))\mu(\beta,Z), A(\beta,Z)\frac{\partial}{\partial\beta}\mu(\beta,Z)$ and $\frac{\partial}{\partial\beta}A(\beta,Z)$ are dominated integrable in a neighborhood of $\beta_0$,
        \item the matrix
        \begin{equation*}
            M=-E\left(A(\beta_0,Z)\frac{\partial}{\partial\beta}\mu(\beta,Z)_{|\beta=\beta_0} \right)
        \end{equation*}
        is invertible,
        \item $E(||A(\beta_0, Z_1)\mu(\beta_0, Z_1)||^{2+\delta})< \infty$ for some $\delta>0.$ 
    \end{enumerate}
    Then, as $n\to\infty$,
    \begin{itemize}
        \item[(a)] the root $\hat{\beta}^B_n$ of $U_n^B(\hat{\beta}_n^B)=0$ exists with a probability tending to 1;
        \item[(b)] $\hat{\beta}_n^B\xrightarrow{p}\beta_0$;
        \item[(c)] $\sqrt{n}(\hat{\beta}_n^B-\hat{\beta}_n)\xrightarrow{d} N_q(0,M^{-1}\Tilde{\Sigma}(M^{-1})^T)$
        \ given the data in probability.
    \end{itemize}
    Mathematically, (c) means
    \begin{equation*}
        \sup_{u\in \mathbb{R}^q}|P(\sqrt{n}(\hat{\beta}_n^B-\hat{\beta}_n) \leq u |(\hat{\theta}_{n,1},Z_1),\dots, (\hat{\theta}_{n,n}, Z_n)) - P(Q\leq u)|\xrightarrow{p}0
    \end{equation*}
    as $n\to\infty,$ where $Q\sim N_q(0,M^{-1}\Tilde{\Sigma}(M^{-1})^T)$ and $\leq$ is taken component-wise.
\end{theorem}

This result demonstrates that the naive bootstrap based on pseudo-observations fails to approximate the correct limit distribution of $\Hat{\beta}_n.$ Instead, a normal distribution with the Huber-White covariance is approximated.
The suggestion of \citet{andersen2003generalised} to bootstrap the pseudo-observations with the aim of variance estimation leads to a biased estimator because Theorem \ref{weakconvbootstrapestimator} implies that the bootstrap variance \eqref{bootstrapvariance} converges to $M^{-1}\Tilde{\Sigma}(M^{-1})^T$ as well (provided it converges at all).

This result may seem surprising but there is an intuitive explanation why the bootstrap fails in this scenario. The bootstrap samples $(\hat{\theta}_{n,1}^B,Z_1^B),\dots, (\hat{\theta}_{n,n}^B, Z_n^B)$ are, by construction, conditionally independent given the data, which is not the case for the original observations. Therefore, there is no possibility to approximate the second order term in the covariance matrix $\Sigma$  because this term arises from the dependence between the pseudo-observations.

Even though the naive bootstrap is invalid for variance estimation, the previous results do not entirely rule out its use for inferential purposes. Wald-type statistics sometimes exhibit a liberal behavior in small samples, and therefore benefit from resampling-based critical values, which often lead to better type-I error control (see e.g.\ \citealt{konietschke2015parametric, pauly2015asymptotic}). 
Noticing that $\Tilde{\Sigma}$ is the limit of the Huber-White estimator $\Hat{\Sigma}^{\text{HW}}_n,$ we can define a bootstrap counterpart of the Wald-statistic

\begin{equation}\label{bootstrapstatistic}
    T_n^B=n(C(\hat{\beta}_n^B-\hat{\beta}_n))^T(C(\Hat{M}_n^{B})^{-1}\Hat{\Sigma}^{\text{HW, B}}_n((\Hat{M}_n^{B})^{-1})^TC^T)^{+}(C(\hat{\beta}_n^B-\hat{\beta}_n)),
\end{equation}
where $\Hat{M}_n^{B}$ and $\Hat{\Sigma}^{\text{HW, B}}_n$ denote the bootstrap counterparts of $\Hat{M}_n$ and $\Hat{\Sigma}^{\text{HW}}_n$, respectively. The following theorem shows that this statistic indeed exhibits the correct limiting distribution---that is, the same limiting distribution as the test statistic $T_n$ under $H_0$.
\begin{theorem}\label{bootstraptest}
    Make the assumptions of Theorem \ref{weakconvbootstrapestimator}. Then, under $H_0$ as well as $H_1,$ we have as $n\to\infty$
    \begin{equation*}
        T_n^B\xrightarrow{d}\chi^2_{\text{rank}(C)}
    \end{equation*}
    given the data in probability.
\end{theorem}

As a consequence of the previous Theorem~\ref{bootstraptest}, we obtain the bootstrap test 
\begin{equation}\label{eq:bootstraptest}
    \varphi^B_{n, \alpha}=\mathbbm{1}(T_n > q^{B}_{n,1-\alpha}),
\end{equation}
where $q^{B}_{n,1-\alpha}$ denotes the conditional $(1-\alpha)$-quantile of $T_n^B.$ The following properties are an immediate consequence of Theorem \ref{bootstraptest}; see  Lemma~1 and Theorem~7 in \citet{janssen2003bootstrap} for details.
\begin{corollary}
    Under the assumptions of Theorem \ref{weakconvbootstrapestimator},
    \begin{enumerate}
        \item[(a)] $\lim_{n\to\infty}E(|\varphi^B_{n, \alpha}-\varphi_{n, \alpha}|)=0$ under $H_0$;
        \item[(b)] $\lim_{n\to\infty}E(\varphi^B_{n, \alpha})=1$ under $H_1$.
    \end{enumerate}
\end{corollary}
Therefore, the bootstrap test shares the same desirable asymptotic properties as the asymptotic test.

With the failure of the bootstrap to approximate the limiting distribution of the parameter estimates in mind, the validity of the bootstrap-based test \eqref{eq:bootstraptest} may be surprising. Two main factors contribute to the validity: the asymptotic pivotality of the Wald-type test statistic and the possibility to consistently estimate the limiting variance of the bootstrapped parameter estimator. This underlines the close connection between the naive bootstrap and the Huber-White estimator in the present context.

\section{Simulations and real data analysis}\label{sec:simulation and data example}
\subsection{Data summary and model}
Our simulation setup is inspired by the Veterans' Administration Lung Cancer study, a randomized clinical trail where two treatment regimes for lung cancer are compared in terms of survival time \citep{prentice1973exponential}. 
This well-known benchmark dataset is available in the R package \texttt{survival} \citep{survival-package}, and contains 137 observations in total, thereof 10 right-censored. In addition to a variable indicating the treatment group (standard or experimental), the survival time (in days) and the censoring status, further covariates are available; in the present analysis, we only use the patients' age (in years) and the tumor cell-type, a factor variable with 4 levels. To keep the model parsimonious, we do not model 
interaction effects.

As estimand of interest, we chose the conditional survival function modeled as 
$S(t_0|Z)=\mu(\beta^T_0 Z)$ 
with the inverse logit, $\mu(x)=\frac{\exp(x)}{1+\exp(x)}$, as a response function, $\beta_0 \in \mathbb{R}^6$, and $Z^T = (1,Z_1, \dots, Z_5)$. We chose $t_0=90,$ therefore our estimand is the probability to survive about three 
months, conditionally on treatment group, tumor cell-type, and age. To ensure identifiability, standard treatment and the ``squamous'' tumor cell-type are treated as reference categories for the respective factors. Therefore, $Z_1$ indicates whether or not a patient receives the experimental treatment, $Z_2,Z_3,Z_4$ are indicators for the second, third, and fourth tumor cell-type, respectively, and $Z_5$ represents the patient's age.

We make the assumption of completely independent censoring, in which case the pseudo-observations can be computed based on the leave-one-out Kaplan-Meier estimators of the event time's survival function.

In addition to testing for the presence of a treatment effect, investigating the influence of the tumor cell-type on the 90-day survival chances is also of interest. Mathematically, if we denote by $\beta_{0,l}$ for $l\in \{1,\dots,6\}$ the $l$th component of $\beta_0$, those hypotheses are
\[H_0^{(1)}:\beta_{0,2}=0 \text{ vs. } H_1^{(1)}:\beta_{0,2}\ne 0,\]
and
\[H_0^{(2)}:\beta_{0,3}=\beta_{0,4}=\beta_{0,5}=0\ \text{ vs. } H_1^{(2)}:\beta_{0,r}\ne 0 \text{ for  at least one } r\in \{3,4,5\}.\]
Both hypotheses can be tested with the help of the above-described Wald-type statistic; suitable hypothesis matrices are
\[C^{(1)}=\begin{pmatrix}
0 & 1 & 0 & 0 & 0 & 0 \\
\end{pmatrix}, ~
C^{(2)}=\begin{pmatrix}
0 & 0 & 1 & 0 & 0 & 0 \\
0 & 0 & 0 & 1 & 0 & 0 \\
0 & 0 & 0 & 0 & 1 & 0
\end{pmatrix}, \]
and $b^{(1)}$ and $ b^{(2)}$ are set to the null vectors of suitable dimensions.
\subsection{Simulation setup}
For our simulations, we strove to mimic the structure of the veteran dataset:
each treatment indicator $Z_1$ is binomially $Bin(1,0.5)$-distributed, the indicators for the four different tumor cell-types are three-dimensional multinomially distributed with equal success probability, and age is modeled as $Z_5\sim N(58, 10.5^2)$; all generated covariates are stochastically independent. Survival times were generated from a Weibull distribution with density 
\[f(t)=\frac{a}{\lambda}\left(\frac{t}{a}\right)^{a-1}e^{-\left(\frac{t}{\lambda}\right)^a}\mathbbm{1}\{t>0\},\]
fixed shape parameter $a=0.85$ (which was obtained by maximum likelihood estimation from the veteran dataset), and individual scale parameters 
\[\lambda=t_0(-\log(\mu(\beta_0^TZ))^{-\frac{1}{a}},\]
which are obtained by solving the survival function $\mu(\beta_0^TZ)=\exp(-(\frac{t_0}{\lambda})^{a})$ for $\lambda.$ We considered the sample sizes $n\in\{80,137,200\},$ of which the second corresponds to the size of our underlying data example. As for censoring, we simulated $C_i \sim U(0,365), U(0,730)$ as well as settings with no censoring, $C_i = \infty$. We summarize this censoring scheme in the vector $\vartheta= (365,730,\infty).$ 
Note that in the case without censoring the pseudo-observation approach reduces to ordinary logistic regression for $A(\beta, Z)=Z$; in any case, we used the function $A(\beta, Z)=\frac{\partial}{\partial\beta}\mu(\beta^TZ)$
which was used in \citet{andersen2003generalised} and originates from generalized estimating equations \citep{liang1986longitudinal}.

The parameter vector was set to $\beta_0=(2.5, \delta_1, \delta_2, 0, 0, -0.04)$ with $\delta_1\in \{0,0.5,1\}$ and $\delta_2\in\{-1,0,1\}.$ For $\delta_1=0$ there is no treatment effect, and $\delta_2=0$ means no difference between the tumor cell-types; otherwise, the respective alternative is implied. All hypotheses were tested at the significance level of $\alpha=0.05$ using the Wald type-statistics based on the corrected and the Huber-White covariance estimators, $\Hat{\Sigma}^{\text{PV}}$ and $\Hat{\Sigma}^{\text{HW}}$, respectively, combined with the critical values from the respective asymptotic chi-squared distributions; we also implemented the bootstrap test described in Section~4. Additionally, we included another variant of the latter two tests where the Huber-White estimator is replaced by the HC3 estimator, which is given by
\begin{equation*}
    \Hat{\Sigma}^{\text{HC3}}_n=\frac{1}{n}\sum_{k=1}^n A(\Hat{\beta}_n, Z_k)A(\Hat{\beta}_n, Z_k)^T(\hat{\theta}_{n,k} -\mu(\Hat{\beta}_n, Z_k))^2(1-d_{kk})^{-2},
\end{equation*}
where $d_{kk}=Z_k^T (\sum_{j=1}^n Z_j Z_j^T)^{-1} Z_k$. This estimator is asymptotically equivalent to the Huber-White estimator, but is recommended in regression models with the sample sizes we consider \citep{long2000using}, and is the default option in the \texttt{sandwich} R package, which implements model-robust covariance estimators for regression models \citep{sandwich_package, sandwich_nonLM}. The R package \texttt{eventglm} \citep{eventglmpackage}, which provides an implementation of the pseudo-observation method, uses the HC3 estimator as a default as well, because if not further specified, \texttt{sandwich} is called for covariance estimation.
The simulations were conducted in the R computing environment, version 4.4.0 \citep{citeR}, each with $n_{\text{sim}}=5000$ simulation runs. The random quantiles $q^{B}_{n,1-\alpha}$ were determined via $B=1000$ bootstrap runs per simulation step.

\subsection{Simulation results}
Table \ref{table:type 1 error like veteran} displays the rejection rates of the asymptotic Wald-type test based on the corrected variance estimator $\varphi^{\text{Corr}}$, the biased Huber-White and HC3 estimators, $\varphi^{\text{HW}}$ and $\varphi^{\text{HC3}}$, respectively,  and two variants of the naive bootstrap tests of Section~\ref{sec:bootstrap}, $\varphi^{B_{\text{HW}}}$ and $\varphi^{B_{\text{HC3}}}$, for testing both null hypotheses $H_0^{(1)}$ and $H_0^{(2)}.$ Here, the critical values of $\varphi^{B_{\text{HW}}}$ are based on the quantiles of \eqref{bootstrapstatistic}, i.e., with the Huber-White estimator of the bootstrap sample for standardization of the bootstrapped test statistic. $\varphi^{B_{\text{HC3}}}$ is constructed in the same manner, but the HC3 estimator of the bootstrap sample is used for standardization instead. All bootstrap-based tests involve the corrected variance estimator for standardization of the test statistic in the original sample. Values that lie within the 95\% binomial confidence interval $[4.4\%,5.6\%]$ are printed in bold. 
The average censoring rates are 30\% for $C_i\sim U(0,730)$ and  46\% for $C_i\sim U(0,365).$

For the single-parameter hypothesis $H_0^{(1)}$, the test based on the corrected variance estimator controls the type-I error reasonably well, especially in large samples, with a slightly liberal behavior in small samples. The test based on the Huber-White estimator behaves very similar but is in general slightly more conservative. This is in line with the findings of \citet{jacobsen2016note} and \citet{overgaard2018estimating} regarding the positive bias of the Huber-White variance estimator which is also apparent in their simulation studies. In case of uncensored observations the corrected and Huber-White estimator are identical, therefore no differences in these scenarios are present. The test based on the HC3 estimator is often too conservative but still keeps the nominal level in many scenarios. The bootstrap tests are often conservative, especially in small samples in the absence of censoring, and fail to reach the desired significance level in many scenarios. However, in some scenarios the bootstrap-based tests are the only tests whose type-I error is within the range of the simulation error.

For the multi-parameter hypothesis $H_0^{(2)}$, the test based on the corrected variance estimator performs well, i.e., the formerly observed liberal behavior is no longer present; instead, the test appears to be somewhat conservative in small sample regimes. For $n=200$ all rejection rates lie inside the 95\% binomial confidence interval. The more conservative behavior of the Huber-White estimator is a bit more distinct in contrast to the single-parameter hypothesis.
The test based on the HC3 estimator is extremely conservative, the nominal level is reached for no scenario. 
Lastly, the behavior of the bootstrap tests is similar as for testing the single-parameter hypothesis; the rejections rates mostly lie in between those of the two asymptotic tests. Also similar to the single-parameter setting, for three scenarios, each with $n=80$ and present censoring, the bootstrap is the only test whose type-I error is within the range of the simulation error.

\begin{table}[H]
\centering
\caption{Rejection rates under the hypotheses $H_0^{(1)}$ and $H_0^{(2)}$. Values in bold lie inside the 95\% binomial confidence interval $[4.4\%, 5.6\%].$}
\vspace{2mm}
\label{table:type 1 error like veteran}
\scriptsize
\begin{tabular}{|rr|r|rrrrr|r|rrrrr|}
\hline
\multicolumn{1}{|l}{} & \multicolumn{1}{l}{} & \multicolumn{1}{|l|}{} & \multicolumn{5}{c|}{$H_0^{(1)}:C^{(1)}\beta_0=0$} & \multicolumn{1}{l|}{} & \multicolumn{5}{c|}{$H_0^{(2)}:C^{(2)}\beta_0=0$} \\
$n$                   & $\vartheta$           & $\delta_2$             & $\varphi^{\text{Corr}}$ & $\varphi^{\text{HW}}$ & $\varphi^{B_\text{HW}}$ & $\varphi^{\text{HC3}}$ & $\varphi^{B_\text{HC3}}$ & $\delta_1$ & $\varphi^{\text{Corr}}$ & $\varphi^{\text{HW}}$ & $\varphi^{B_\text{HW}}$ & $\varphi^{\text{HC3}}$ & $\varphi^{B_\text{HC3}}$ \\ \hline
80 & $\infty$ & 0.0 & 5.7 & 5.7 & 1.3 & 3.7 & 1.2 & 0.0 & 3.9 & 3.9 & 0.1 & 1.7 & 0.1 \\ 
137 & $\infty$ & 0.0 & $\mathbf{5.3}$ & $\mathbf{5.3}$ & 3.8 & $\mathbf{4.5}$ & 3.9 & 0.0 & $\mathbf{4.9}$ & $\mathbf{4.9}$ & 3.7 & 3.0 & 3.6 \\ 
200 & $\infty$ & 0.0 & 5.8 & 5.8 & $\mathbf{4.9}$ & $\mathbf{4.7}$ & 4.2 & 0.0 & $\mathbf{4.7}$ & $\mathbf{4.7}$ & 4.2 & 4.0 & $\mathbf{4.6}$ \\ 
80 & 730 & 0.0 & 6.0 & 5.9 & $\mathbf{4.6}$ & 3.9 & 4.3 & 0.0 & 4.2 & 3.9 & 4.2 & 1.5 & $\mathbf{4.4}$ \\ 
137 & 730 & 0.0 & $\mathbf{5.5}$ & $\mathbf{5.4}$ & 3.7 & 4.3 & 3.9 & 0.0 & 4.3 & 4.2 & 3.8 & 2.7 & 3.7 \\ 
200 & 730 & 0.0 & $\mathbf{5.1}$ & $\mathbf{5.1}$ & 4.0 & $\mathbf{4.7}$ & $\mathbf{4.5}$ & 0.0 & 4.3 & 4.2 & 4.2 & 3.7 & $\mathbf{4.5}$ \\ 
80 & 365 & 0.0 & $\mathbf{5.4}$ & $\mathbf{5.2}$ & $\mathbf{5.1}$ & 3.7 & $\mathbf{5.5}$ & 0.0 & 3.4 & 2.9 & $\mathbf{4.6}$ & 1.1 & $\mathbf{4.4}$ \\ 
137 & 365 & 0.0 & $\mathbf{5.4}$ & $\mathbf{5.3}$ & 4.0 & $\mathbf{4.6}$ & 4.3 & 0.0 & $\mathbf{4.4}$ & 4.2 & $\mathbf{4.4}$ & 2.4 & $\mathbf{4.4}$ \\
200 & 365 & 0.0 & $\mathbf{5.3}$ & $\mathbf{5.3}$ & $\mathbf{4.4}$ & 4.2 & 4.2 & 0.0 & $\mathbf{4.5}$ & 4.3 & $\mathbf{4.5}$ & 3.7 & $\mathbf{4.7}$ \\ \hline  
80 & $\infty$ & -1.0 & 6.6 & 6.6 & 1.2 & 4.0 & 1.1 & 0.5 & 4.2 & 4.2 & 0.0 & 2.2 & 0.1 \\ 
137 & $\infty$ & -1.0 & 5.7 & 5.7 & 3.9 & $\mathbf{4.4}$ & 3.6 & 0.5 & $\mathbf{4.9}$ & $\mathbf{4.9}$ & 3.4 & 2.9 & 2.8 \\ 
200 & $\infty$ & -1.0 & 5.7 & 5.7 & 4.3 & $\mathbf{5.2}$ & $\mathbf{4.9}$ & 0.5 & $\mathbf{4.7}$ & $\mathbf{4.7}$ & 4.3 & 3.6 & 4.0 \\ 
80 & 730 & -1.0 & 6.3 & 6.1 & $\mathbf{5.0}$ & 4.0 & $\mathbf{4.8}$ & 0.5 & 3.4 & 3.2 & 2.8 & 1.5 & 3.5 \\ 
137 & 730 & -1.0 & $\mathbf{5.0}$ & $\mathbf{5.0}$ & 3.7 & $\mathbf{4.6}$ & 4.1 & 0.5 & $\mathbf{4.9}$ & $\mathbf{4.7}$ & 3.7 & 2.9 & 3.4 \\ 
200 & 730 & -1.0 & $\mathbf{5.4}$ & $\mathbf{5.4}$ & $\mathbf{4.4}$ & $\mathbf{4.7}$ & 4.2 & 0.5 & $\mathbf{4.5}$ & $\mathbf{4.4}$ & 4.1 & 3.5 & 3.8 \\ 
80 & 365 & -1.0 & 6.3 & 6.0 & 6.1 & 3.6 & 5.7 & 0.5 & 3.2 & 2.8 & 4.2 & 1.2 & $\mathbf{4.4}$ \\ 
137 & 365 & -1.0 & $\mathbf{5.5}$ & $\mathbf{5.3}$ & 4.0 & 4.2 & 4.0 & 0.5 & 3.7 & 3.4 & 3.5 & 2.6 & 3.9 \\ 
200 & 365 & -1.0 & $\mathbf{5.5}$ & $\mathbf{5.3}$ & 4.1 & $\mathbf{4.5}$ & 4.3 & 0.5 & $\mathbf{4.9}$ & $\mathbf{4.7}$ & $\mathbf{4.5}$ & 3.1 & 4.2 \\ \hline
80 & $\infty$ & 1.0 & 5.9 & 5.9 & 0.9 & 3.7 & 0.8 & 1.0 & $\mathbf{4.5}$ & $\mathbf{4.5}$ & 0.1 & 2.2 & 0.1 \\ 
137 & $\infty$ & 1.0 & $\mathbf{5.2}$ & $\mathbf{5.2}$ & 3.3 & 4.1 & 3.3 & 1.0 & $\mathbf{4.9}$ & $\mathbf{4.9}$ & 1.8 & 2.7 & 1.9 \\ 
200 & $\infty$ & 1.0 & $\mathbf{5.2}$ & $\mathbf{5.2}$ & 4.1 & $\mathbf{4.4}$ & 3.8 & 1.0 & $\mathbf{4.9}$ & $\mathbf{4.9}$ & 3.4 & 3.7 & 3.6 \\ 
80 & 730 & 1.0 & 5.9 & 5.8 & $\mathbf{4.5}$ & 3.2 & 3.9 & 1.0 & 3.9 & 3.5 & 2.7 & 0.9 & 3.1 \\ 
137 & 730 & 1.0 & $\mathbf{5.3}$ & $\mathbf{5.2}$ & 3.6 & 4.1 & 3.7 & 1.0 & 3.6 & 3.4 & 2.4 & 2.3 & 2.8 \\ 
200 & 730 & 1.0 & $\mathbf{5.4}$ & $\mathbf{5.4}$ & 4.2 & $\mathbf{4.8}$ & $\mathbf{4.4}$ & 1.0 & $\mathbf{4.6}$ & $\mathbf{4.5}$ & 3.7 & 3.1 & 3.5 \\ 
80 & 365 & 1.0 & 5.7 & $\mathbf{5.5}$ & 5.7 & 4.1 & 6.0 & 1.0 & 3.6 & 2.9 & 3.8 & 1.1 & 3.4 \\ 
137 & 365 & 1.0 & $\mathbf{5.6}$ & $\mathbf{5.4}$ & 3.8 & $\mathbf{4.4}$ & 4.3 & 1.0 & 4.0 & 3.7 & 3.1 & 2.3 & 3.6 \\ 
200 & 365 & 1.0 & $\mathbf{5.5}$ & $\mathbf{5.5}$ & 4.1 & $\mathbf{4.8}$ & 4.2 & 1.0 & $\mathbf{4.5}$ & 4.3 & 3.9 & 3.3 & 3.8 \\             \hline
\end{tabular}
\end{table}

Regarding the power behavior of the tests under the alternative hypotheses, all sizes point to a similar picture.
As a consequence,  we only present the results for the multi-parameter hypothesis $H_0^{(2)}$ with the true effect $\delta_2=-1$ in Table~\ref{table:power like veteran celltype-1}. Results for $\delta_2=1$ as well as $\delta_1\in\{0.5,1\}$ are 
available in Appendix \ref{appendix:C add simulation}. As expected, the test based on the corrected variance estimator exhibits a greater power than the tests based on the Huber-White and HC3 estimators; this difference is most pronounced in small samples, up to a difference in power as big as $11.4$ percentage points for the HC3 estimator. For the Huber-White estimator the differences are smaller, which is in line with the behavior under the null hypothesis. The power of the bootstrap test mostly lies in between the power for the asymptotic tests.
In some scenarios, however, they have the highest power. Furthermore, even when the power of the test based on the corrected variance estimator is higher, that difference is only small for the settings with high censoring proportions. Therefore, the conservativeness of the bootstrap tests under the null hypothesis does not always lead to a decrease in power. Between the two bootstrap tests, no systematic differences are noticeable. With increasing sample sizes, the power of all tests increases and differences between the tests diminish. Furthermore, the power of all tests decreases with increasing censoring which was to be expected.

\begin{table}[H]
\centering
\caption{Power of the tests under $H_1^{(2)}:C^{(2)}\beta_0\ne 0$, the true effect is $\delta_2=-1.$ The test with the highest power for each scenario is printed in bold.}
\vspace{2mm}
\label{table:power like veteran celltype-1}
\scriptsize
\begin{tabular}{|rr|r|rrrrr|}
\hline
$n$                   & $\vartheta$                & $\delta_1$            & $\varphi^{\text{Corr}}$ & $\varphi^{\text{HW}}$ & $\varphi^{B_{\text{HW}}}$ & $\varphi^{\text{HC3}}$ & $\varphi^{B_{\text{HC3}}}$ \\ \hline
$80$ & $\infty$ & $0.0$ & $\mathbf{24.9}$ & $\mathbf{24.9}$ & $1.1$ & $13.9$ & $0.9$ \\ 
$137$ & $\infty$ & $0.0$ & $\mathbf{46.1}$ & $\mathbf{46.1}$ & $41.8$ & $39.5$ & $41.6$ \\ 
$200$ & $\infty$ & $0.0$ & $\mathbf{66.9}$ & $\mathbf{66.9}$ & $65.8$ & $62.4$ & $65.2$ \\ 
$80$ & $730$ & $0.0$ & $20.5$ & $19.6$ & $20.4$ & $9.8$ & $\mathbf{20.8}$ \\ 
$137$ & $730$ & $0.0$ & $\mathbf{44.2}$ & $43.8$ & $43.3$ & $34.8$ & $40.9$ \\ 
$200$ & $730$ & $0.0$ & $\mathbf{63.3}$ & $63.0$ & $63.1$ & $57.1$ & $61.3$ \\ 
$80$ & $365$ & $0.0$ & $16.8$ & $14.9$ & $20.0$ & $7.0$ & $\mathbf{21.1}$ \\ 
$137$ & $365$ & $0.0$ & $38.8$ & $38.0$ & $\mathbf{39.3}$ & $30.8$ & $39.2$ \\ 
$200$ & $365$ & $0.0$ & $57.5$ & $57.0$ & $57.9$ & $52.5$ & $\mathbf{58.3}$ \\ 
$80$ & $\infty$ & $0.5$ & $\mathbf{25.5}$ & $\mathbf{25.5}$ & $0.8$ & $15.3$ & $1.0$ \\ 
$137$ & $\infty$ & $0.5$ & $\mathbf{49.0}$ & $\mathbf{49.0}$ & $44.0$ & $41.5$ & $42.9$ \\ 
$200$ & $\infty$ & $0.5$ & $\mathbf{67.5}$ & $\mathbf{67.5}$ & $66.0$ & $63.0$ & $65.4$ \\ 
$80$ & $730$ & $0.5$ & $\mathbf{22.1}$ & $21.2$ & $21.4$ & $10.6$ & $20.9$ \\ 
$137$ & $730$ & $0.5$ & $\mathbf{43.9}$ & $43.4$ & $42.1$ & $36.7$ & $41.8$ \\ 
$200$ & $730$ & $0.5$ & $\mathbf{63.8}$ & $63.5$ & $62.8$ & $59.8$ & $63.0$ \\ 
$80$ & $365$ & $0.5$ & $18.0$ & $16.1$ & $20.4$ & $7.3$ & $\mathbf{20.7}$ \\ 
$137$ & $365$ & $0.5$ & $40.7$ & $39.7$ & $\mathbf{41.0}$ & $32.6$ & $39.6$ \\ 
$200$ & $365$ & $0.5$ & $\mathbf{59.4}$ & $58.9$ & $59.0$ & $54.6$ & $58.5$ \\ 
$80$ & $\infty$ & $1.0$ & $\mathbf{24.4}$ & $\mathbf{24.4}$ & $0.4$ & $13.0$ & $0.2$ \\ 
$137$ & $\infty$ & $1.0$ & $\mathbf{47.5}$ & $\mathbf{47.5}$ & $38.7$ & $39.7$ & $37.1$ \\ 
$200$ & $\infty$ & $1.0$ & $\mathbf{64.2}$ & $\mathbf{64.2}$ & $60.9$ & $61.5$ & $62.9$ \\ 
$80$ & $730$ & $1.0$ & $\mathbf{21.2}$ & $20.4$ & $19.5$ & $9.7$ & $18.7$ \\ 
$137$ & $730$ & $1.0$ & $\mathbf{43.2}$ & $42.9$ & $39.3$ & $35.3$ & $38.4$ \\ 
$200$ & $730$ & $1.0$ & $59.8$ & $59.6$ & $57.8$ & $58.2$ & $\mathbf{60.3}$ \\ 
$80$ & $365$ & $1.0$ & $15.9$ & $14.4$ & $17.5$ & $6.9$ & $\mathbf{17.6}$ \\ 
$137$ & $365$ & $1.0$ & $\mathbf{39.5}$ & $38.2$ & $37.3$ & $31.0$ & $36.4$ \\ 
$200$ & $365$ & $1.0$ & $\mathbf{58.5}$ & $58.0$ & $57.0$ & $53.3$ & $56.5$ \\   \hline
\end{tabular}
\end{table}
Based on these practical and our theoretical findings, we draw the following recommendations:
\begin{enumerate}
\item[(i)] For testing single parameter hypotheses and therefore also for confidence intervals for single elements of the parameter vector, the naive bootstrap can be applied in small samples, with the constraint that it should not be used in the case of uncensored observations. If strict type-I error control is necessary, tests based on the HC3 estimator can be used; however, this can lead to a severe loss in power. In larger samples, tests based on corrected variance estimator should be used. This is in line with the suggestions of \citet{jacobsen2016note} as well as \citet{overgaard2018estimating}, which are similar but based on simulations of lower dimensional models.

\item[(ii)]
For testing multi-parameter hypotheses, the asymptotic test based on the corrected variance estimator performs well, even in small samples; 
it is our preferred choice.
The naive bootstrap provides an attractive alternative if a considerable amount of censoring is observed. 

\item[(iii)] If the corrected estimator is not available, tests based on the Huber-White estimator are recommended, because the loss in power is less severe compared to tests based on the HC3 estimator.
\end{enumerate}
Further simulation results for a model with an interaction effect can be found in Appendix \ref{appendix:C add simulation}. For this model greater differences in power between the tests $\varphi^{\text{Corr}}$ and $\varphi^{\text{HW}}$ were observed.

\subsection{Data Analysis}
Table~\ref{table: data example} summarizes the point estimates as well as the test results for the null hypotheses of no treatment and no cell-type effect, respectively, for the veteran dataset. Because the censoring proportion is relatively low---only 7.3\%---we have not included the bootstrap tests, as they are expected to be rather conservative.

\begin{table}[H]
\centering
\caption{Pseudo-observation regression estimates for 90 day survival, including standard errors and Wald-type tests calculated with the corrected variance estimator $\Hat{\Sigma}^{\text{PV}}_n$, the Huber-White estimator $\Hat{\Sigma}^{\text{HW}}_n$ and the HC3 estimator $\Hat{\Sigma}^{\text{HC3}}_n$. trt2 denotes the experimental treatment group. The tumor cell-types small, adeno and large, are denoted by celltypesmallcell, celltypeadeno and celltypelarge. $t_n(k)$ denote the test scores where $k$ is the rank of the corresponding hypothesis matrix $C.$}
\vspace{2mm}
\label{table:veteran data example}
\scriptsize
\begin{threeparttable}
\begin{tabular}{l|c|c|c|c}
\hline
 &  & $\Hat{\Sigma}^{\text{PV}}_n$ & $\Hat{\Sigma}^{\text{HW}}_n$ & $\Hat{\Sigma}^{\text{HC3}}_n$ \\
Variable & Coefficient & Standard error & Standard error & Standard error \\
\hline
(Intercept) & 1.542 & 1.164 & 1.165 & 1.243 \\
trt2 & -0.772 & $0.407$ & 0.407 & $0.429$ \\
celltypesmallcell & -1.640 & $0.527$ & 0.527 & $0.551$ \\
celltypeadeno & -1.186 & $0.546$ & 0.547 & $0.577$ \\
celltypelarge & 0.318 & 0.581 & 0.581 & 0.617 \\
age & -0.009 & 0.019 &0.019  & 0.021 \\
\hline
$H_0^{(1)}:C^{(1)}\beta_0=0$ &  & $t_n(1)=3.597, p=0.058$ & $t_n(1)=3.587, p=0.058$ & $t_n(1)=3.201, p=0.072$  \\
$H_0^{(2)}:C^{(2)}\beta_0=0$ &  & $t_n(3)=18.098, p<0.001$ & $t_n(3)=18.068, p<0.001$ & $t_n(3)=16.430, p<0.001$  \\
\hline
\end{tabular}

\end{threeparttable}
\label{table: data example}
\end{table}
The results indicate a strong effect of the tumor cell-type on patients' survival, but no effect of the experimental treatment when the standard significance level of $\alpha=0.05$ is used. The values for the test statistics as well as the standard errors are almost indistinguishable between the corrected and Huber-White variance estimator; and the differences are also small if those are compared to the HC3 estimator. Recalling the low censoring rate in the dataset, this is not surprising. Furthermore, the standard errors are smaller for the corrected variance estimator, leading to larger values of the Wald-type statistics. These observations are also in line with our simulation results.

\section{Discussion and recommendation}\label{sec:discussion}

The Huber-White and HC3 covariance estimators in the pseudo-observation regression approach have been used to date, in both methodological and applied work.
One factor that may contribute to this is their wide availability in statistical software.

In this paper, we have verified that these estimators are biased and we observed in our simulations that their usage can result in conservative tests.
Thus, in general, the corrected covariance estimator seems a more recommendable choice.
However, one obstacle for its use is that it is currently only available for a few estimands. This issue remains, as the structure of the corrected variance estimator requires a separate implementation for each estimand.

We have also verified that the naive bootstrap is not available for an unbiased covariance estimation although it can be used for hypothesis testing if combined with an appropriate studentization.
As our simulation results suggest, the power of tests for general linear hypotheses benefits from our proposed methods, i.e., the usage of the corrected covariance estimator and the bootstrap. On the other hand, we observed that the use of the HC3 modification of the Huber-White estimator leads to an accurate type-I error control at the cost of losing power. 
Based on the investigations of covariances in \citet{jacobsen2016note, overgaard2018estimating, overgaard2025comparison}, we also expect that our simulation-based findings generalize to models based on link functions other than the logit function and to other estimands.

In future research, we will also investigate other means of bootstrapping which have been suggested by \citet{overgaard2018estimating,parner2023regression};
this concerns randomly drawing with replacement from the original observations and then re-solving the estimating equations as well as combining this bootstrap with 
infinitesimal jackknife pseudo-observations. Furthermore, we aim to provide ready-to-use software implementations for these approaches in the future.

\section*{Acknowledgments}

We would like to thank our previous affiliations TU Dortmund University and Research Center Trustworthy Data Science and Security (Simon Mack, Dennis Dobler) and Otto von Guericke University Magdeburg (Simon Mack) where part of the work was done. The authors gratefully acknowledge the computing time
provided on the Linux HPC cluster at TU
Dortmund University (LiDO3), partially funded in the course of the
Large-Scale Equipment Initiative by the Deutsche
Forschungsgemeinschaft (DFG, German Research Foundation) as
project 271512359.
Additionally, we thank Sven Schulz-Niethammer whose preliminary simulation study gave good guidance for the present one.
Finally, special thanks go to our late friend and colleague Marc Ditzhaus for helpful discussions and guidance of Simon Mack in an early phase of the project.

\bibliographystyle{abbrvnat}
\bibliography{literature}
\newpage

\appendix

\section*{Appendix}\label{appendix:A uniform LLN}

The appendices are organized as follows: Appendix A gives an introduction to U-and V-statistics in Banach spaces, for which a uniform law of large numbers is proposed. Appendix B contains the proofs of the results from the main part, as well as additional mathematical results that are necessary for these proofs. In Appendix C we present the remaining simulation results from the simulation study of the main paper, as well as results for an additional simulation setup.

\section{U- and V-statistics in Banach spaces}
Our proofs heavily rely on results for U- and V-statistics which are not necessarily real-valued. Therefore, we give a short introduction in a very general setting, and propose a uniform law of large numbers. A very detailed treatment of such statistics can be found in \citet{borovskikh1996ustatistics}. Let $B$ be a separable Banach space with norm $||\cdot||_B$ and $X_1,\dots,X_n$ independent random variables on a measurable space $(\mathcal{X}, \mathcal{A})$ with identical distribution $P.$ Consider a symmetric function $\Phi:\mathcal{X}^m\to B$ of $m$ variables.
A $B$-valued U-Statistic is defined as follows:
\[U_n^{(m)}(\Phi)=\binom{n}{m}^{-1}\sum_{1\leq i_1<\dots<i_m\leq n}\Phi(X_{i_1},\dots,X_{i_m}).\]
The number $m$ is called the degree of the U-statistic. If the \textit{kernel} $\Phi$ is Bochner-integrable w.r.t. $P$, $U_n^{(m)}(\Phi)$ is an unbiased estimator of $E(\Phi(X_1,\dots,X_m)).$ Additionally, we define the corresponding V-Statistic
\[V_n^{(m)}(\Phi)=\frac{1}{n^m}\sum_{i_1=1}^n\dots\sum_{i_m=1}^n\Phi(X_{i_1},\dots,X_{i_m}).\]
Let $(X_1^{\ast},\dots,X_N^{\ast})$ be a bootstrap sample of size $n$ which is obtained by sampling $n$ times with replacement from $(X_1,\dots,X_n).$ Then the bootstrap U-statistic is defined as
\[U_n^{(m)\ast}(\Phi)=\binom{n}{m}^{-1}\sum_{1\leq i_1\ne\dots\ne i_m\leq n}\Phi(X^{\ast}_{i_1},\dots,X^{\ast}_{i_m}).\]
The bootstrap V-statistic $V_n^{(m)\ast}(\Phi)$ is defined accordingly. For applications, the kernel is also allowed to depend on a fixed but unknown parameter. As we show subsequently, laws of large numbers can be obtained uniformly in this parameter under mild regularity conditions.
\begin{assumption}\label{kernelcontandmeasure}
Let $\Theta$ be a compact subset of separable metric space and $\Phi_{\theta}=\Phi(\theta, x_1,\dots,x_m)$ a $B$-valued function on $\Theta \times \mathcal{X}^m$ such that
    \begin{enumerate}
        \item[(i)]for each $(x_1,\dots,x_m)\in \mathcal{X}^m$, $\theta\mapsto\Phi(\theta, x_1,\dots,x_m)$ is continuous on $\Theta,$
        \item[(ii)] for each $\theta \in \Theta$, $(x_1,\dots,x_m)\mapsto \Phi(\theta, x_1,\dots,x_m)$ is a measurable and symmetric function on $\mathcal{X}^m.$
    \end{enumerate}
\end{assumption}

\begin{assumption}\label{uniformmomentcond}
    Suppose that
    \begin{enumerate}
        \item[(1)] $$E(\sup_{\theta\in \Theta}||\Phi(\theta, X_1,\dots,X_m)||_B)<\infty,$$
        \item[(2)] for each possible combination of integers $i_1,\dots,i_m$
        $$E(\sup_{\theta\in \Theta}||\Phi(\theta, X_{i_1},\dots,X_{i_m})||^{\#\{i_1,\dots,i_m\}/m}_B)<\infty$$
        where $\#\{i_1,\dots,i_m\}$ denotes the cardinality of the set $\{i_1,\dots,i_m\}.$
        \item[(3)] The Banach space $B$ is of Type 2; see the subsequent remark for details.
    \end{enumerate}
\end{assumption}
\begin{remark}
     A Banach Space is of Type 2 if there exists a constant $b$ such that for every i.i.d.\ vector of $B$-valued random elements $(\xi_1,\dots,\xi_n)$ with $E(\xi_1)=0, E(||\xi_1||_B^2)<\infty$  and all $n\geq 1$ the inequality
\[E(||\sum_{i=1}^n\xi_i||^2_B)\leq b\sum_{i=1}^nE(||\xi_i||_B^2)\]
holds. Every Hilbert space is of Type 2.
\end{remark}

\begin{theorem}\label{Uniform LLN U and V}
Suppose that Assumption \ref{kernelcontandmeasure} holds.
    As $n\to\infty$,
    \begin{enumerate}
        \item[(a)] and, if Assumption \ref{uniformmomentcond} (1) holds, then
        $$\sup_{\theta\in\Theta}||U_n^{(m)}(\Phi_{\theta})-E(\Phi(\theta, X_1,\dots,X_m))||_B \xrightarrow{\text{a.s.}} 0;$$ 
        \item[(b)] and, if Assumption \ref{uniformmomentcond} (2) holds, then $$\sup_{\theta\in\Theta}||V_n^{(m)}(\Phi_{\theta})-E(\Phi(\theta, X_1,\dots,X_m))||_B \xrightarrow{\text{a.s.}} 0;$$ 
        \item[(c)] and, if Assumptions \ref{uniformmomentcond} (2) and (3) hold, then  $$\sup_{\theta\in\Theta}||U_n^{(m)\ast}(\Phi_{\theta})-E(\Phi(\theta, X_1,\dots,X_m))||_B \xrightarrow{p} 0,$$ as well as $$P(||U_n^{(m)\ast}(\Phi_{\theta})-E(\Phi(\theta, X_1,\dots,X_m))||_B >\varepsilon | X_1,\dots,X_n)\xrightarrow{\text{a.s.}} 0$$ for all $\varepsilon
        >0.$ These results hold for the bootstrap V-statistic $V_n^{(m)\ast}(\Phi_{\theta})$ as well.
    \end{enumerate}
\end{theorem}

\section{Proofs and auxiliary mathematical results}\label{appendix:B proofs}

First of all, we state a lemma that relates conditional and unconditional convergence in probability.

\begin{lemma}\label{pconvergenceequiv}
Let $X_n:\Omega\to\mathbb{D}$ be a sequence of random variables on a probability space $(\Omega, \mathcal{A}, P)$ with values in a separable metric space $(\mathbb{D}, d_{\mathbb{D}}).$ Additionally let $\mathcal{B}_n\subset\mathcal{A}$ be a sequence of sub $\sigma$-fields.
    Then as $n\to \infty,$ $d_{\mathbb{D}}(X_n, X)\xrightarrow{p}0$ for a $\mathbb{D}$ valued random variable $X$ if and only if $P(d_{\mathbb{D}}(X_n, X)>\varepsilon |\mathcal{B}_n)\xrightarrow{p}0$ for every $\varepsilon>0.$
\end{lemma}
    A proof of this result (for the Skorokhod space) was already given in the supplement of \citet{dobler2019confidence}, and the subsequent proof is a consequence of merely adapting the notation.
\begin{proof}[Proof of Lemma \ref{pconvergenceequiv}]
Define the function $Q_n(\varepsilon)=P(d_{\mathbb{D}}(X_n, X)>\varepsilon |\mathcal{B}_n)$ and suppose that $d_{\mathbb{D}}(X_n, X)\xrightarrow{p}0$ as $n\to\infty.$ Then we have by Markov's inequality for any $c>0,$
\begin{equation*}
    \begin{split}
        P(Q_n(\varepsilon)>c)&\leq\frac{E(Q_n(\varepsilon))}{c}\\
        &=\frac{P(d_{\mathbb{D}}(X_n, X)>\varepsilon)}{c}\to0.
    \end{split}
\end{equation*}
Now suppose $Q_n(\varepsilon)\xrightarrow{p}0$ as $n\to\infty$. Because $Q_n(\varepsilon)\leq1,$ and due to the subsequence criterion, we can apply the dominated convergence theorem to conclude
\[E(Q_n(\varepsilon))=P(d_{\mathbb{D}}(X_n, X)>\varepsilon)\to0.\]
\end{proof}

We can now prove the uniform law of large numbers for U- and V-statistics.

\begin{proof}[Proof of Theorem \ref{Uniform LLN U and V}]
    The technique of proof for these kind of uniform laws is well known, and we follow the structure of the proofs in \citet{spencer2024strong}. Define $g(\theta)=E(\Phi(\theta, X_1,\dots,X_m))$ and note that $g(\theta)$ is continuous by the dominated convergence theorem if Assumption \ref{uniformmomentcond} (1) or (2) is fulfilled. Let $D_r(\theta_0)$ be the open ball with radius $r$, centered at $\theta_0\in\Theta$ (with respect to the metric on $\Theta$). For $\theta\in\Theta$, $(x_1,\dots,x_m)\in \mathcal{X}^m$ and $r>0$ define the function
    \[q(\theta, x_1,\dots,x_m,r):= \sup_{\theta'\in D_r(\theta)}\left\Vert (\Phi(\theta, x_1,\dots,x_m)-g(\theta))-(\Phi(\theta', x_1,\dots,x_m)-g(\theta'))  \right\Vert_B  \]
    This function is still symmetric in the arguments $(x_1,\dots,x_m).$ Additionally by continuity and because $\Theta$ is compact
    \begin{equation}\label{q-finite}
        q(\theta, x_1,\dots,x_m,r) \leq 2\sup_{\theta\in\Theta}||\Phi(\theta, x_1,\dots,x_m)||_B+2\sup_{\theta\in\Theta}||g(\theta)||_B<\infty.
    \end{equation}
    Using the dominated convergence theorem again, it follows that
    \begin{equation*}
        \lim_{r\to0}E(q(\theta, x_1,\dots,x_m,r))=0
    \end{equation*} for all $\theta \in \Theta.$ Therefore and by compactness for every $\varepsilon>0$ exists $L\in\mathbb{N},$ $\theta_1,\dots, \theta_L$ and $r_1,\dots,r_L$ such that $\Theta=\bigcup_{j=1}^L D_{r_j}(\theta_j)$ and $E(q(\theta_j, x_1,\dots,x_m,r_j))<\varepsilon.$ Now
    \begin{equation}\label{q-convergence}
        \binom{n}{m}^{-1}\sum_{1\leq i_1<\dots<i_m\leq n}q(\theta_j, X_{i_1},\dots,X_{i_m},r_j)\xrightarrow{\text{a.s.}} E(q(\theta_j, X_1,\dots,X_m,r_j))<\varepsilon
    \end{equation}
    by the strong law of large numbers for Banach Space valued U-statistics (Theorem 3.1.1 of \citet{borovskikh1996ustatistics}). Similarly
    \begin{equation}\label{psi-convergence}
        \binom{n}{m}^{-1}\sum_{1\leq i_1<\dots<i_m\leq n}\Phi(\theta_j, X_{i_1},\dots,X_{i_m}) \xrightarrow{\text{a.s.}} g(\theta_j).
    \end{equation}
    Therefore for every $\theta \in \Theta$, if we choose $j$ such that $\theta\in D_{r_j}(\theta_j),$ we have by the triangle inequality
    \begin{align*}   
        \left\Vert U_n^{(m)}(\Phi_{\theta})-g(\theta)\right\Vert_B
        \leq& \binom{n}{m}^{-1}\sum_{1\leq i_1<\dots<i_m\leq n}q(\theta_j, X_{i_1},\dots,X_{i_m},r_j)\\ &+ \left\Vert \binom{n}{m}^{-1}\sum_{1\leq i_1<\dots<i_m\leq n}\Phi(\theta_j, X_{i_1},\dots,X_{i_m})-g(\theta_j) \right\Vert_B\\
        :=&S_{n,j}.
    \end{align*}
    By \eqref{q-convergence} and \eqref{psi-convergence} we have $S_{n,j}\xrightarrow{\text{a.s.}} E(q(\theta_j, X_1,\dots,X_m,r_j))<\varepsilon.$ Therefore,
    \begin{align*}
        \sup_{\theta\in\Theta}||U_n^{(m)}(\Phi_{\theta})&-E(\Phi(\theta, X_1,\dots,X_m))||_B\\
        &\leq \max_{1\leq j\leq L} S_{n,j} \xrightarrow{\text{a.s.}} \max_{1\leq j\leq L} E(q(\theta_j, X_1,\dots,X_m,r_j))<\varepsilon
    \end{align*}
    This proves (a), since $\varepsilon>0$ was arbitrary.

    The V-statistic $V_n^{(m)}(\Phi_{\theta})$ can be written as a linear combination of U-statistics of which the leading term is $U_n^{(m)}(\Phi_{\theta})$ and all the other terms are of lower order, see Equation (1.3.3) in \citet{borovskikh1996ustatistics}. For the leading term, we can apply (a). Due to Assumption \ref{uniformmomentcond} (2) we can apply Corollary 3.4.2. of \citet{borovskikh1996ustatistics} to the $m-1$ lower order terms, which proves (b).

    Similar to (a) but by applying the bootstrap law of large numbers for U-Statistics (Theorem 13.3.1 in \citealt{borovskikh1996ustatistics}), (c) holds almost surely, and therefore in probability conditional on the data. Because conditional convergence in probability is equivalent to unconditional convergence in probability (Lemma \ref{pconvergenceequiv}), (c) also holds unconditionally. As in the proof of (b) we can decompose $V_n^{(m)\ast}(\Phi_{\theta})$ into a linear combination of bootstrap U-statistics, with leading term $U_n^{(m)\ast}(\Phi_{\theta}).$ Applying the already proven uniform law to each summand, the result for bootstrap V-statistics follows.
\end{proof}

The convergence in distribution of the Wald-type statistic is based on the asymptotic normality of the parameter estimates. We state this convergence result (Theorem 3.4 of \citealt{overgaard2017}) for the sake of completeness, and in order to elaborate on some details of the proof.
\begin{theorem}\label{overgaard main variant}
    Consider estimating equations
    \begin{equation*}\label{estimatingequation}
         U_n(\beta)=\sum_{k=1}^n A(\beta, Z_k)(\hat{\theta}_{n,k} -\mu(\beta, Z_k)),
     \end{equation*} 
     and make the assumptions of Theorem \ref{weakconvbootstrapestimator} with the modification that $E(||A(\beta_0, Z_1)\mu(\beta_0, Z_1)||^{2})< \infty$ is sufficient.
    Then, as $n\to\infty$,
    \begin{itemize}
        \item[(a)] the root $\hat{\beta}_n$ of $U_n(\hat{\beta}_n)=0$ exists with a probability tending to 1;
        \item[(b)] $\hat{\beta}_n\xrightarrow{p}\beta_0$;
        \item[(c)] $\sqrt{n}(\hat{\beta}_n-\beta_0)\xrightarrow{d} N_q(0,M^{-1}{\Sigma}(M^{-1})^T).$
    \end{itemize}
        
\end{theorem}

\begin{proof}[Proof of Theorem \ref{overgaard main variant}]
 We aim to apply Theorem 2.11 of \citet{jacod2018review}. The asymptotic normality of the estimating function was already established by \citet{overgaard2017} and we have nothing to add here. It remains to show, that
    \begin{equation*}\label{averagederivativematrix}
        \begin{split}
        \frac{\partial}{\partial \beta}(\frac{1}{n}U_n(\beta))= & \frac{1}{n}\sum_{k=1}^n \frac{\partial}{\partial\beta}(A(\beta, Z_k))(\hat{\theta}_{n,k} -\mu(\beta, Z_k)) \\
        & - \frac{1}{n} \sum_{k=1}^n A(\beta, Z_k)\frac{\partial}{\partial\beta}\mu(\beta, Z_k)
    \end{split}
    \end{equation*}
    converges in probability to $M=M(\beta_0)$ uniformly in a neighborhood $G$ of $\beta_0.$ It is no restriction to assume that $G$ is the closed $\varepsilon$-ball $\overline{B}_{\varepsilon}(\beta_0)$ centered at $\beta_0,$ because it always contains this ball if $\varepsilon$ is chosen small enough. Similarly, we can assume that the neighborhood on which the dominated integrability conditions hold, coincides with $G.$
     We first show, that uniform convergence in probability holds for
        \begin{equation}\label{derivativeUstatistic}
        \begin{split}
        \frac{\partial}{\partial \beta}(\frac{1}{n}U_n^{\ast}(\beta))= & \frac{1}{n}\sum_{k=1}^n \frac{\partial}{\partial\beta}(A(\beta, Z_k))(\hat{\theta}^{\ast}_{n,k} -\mu(\beta, Z_k)) \\
        & - \frac{1}{n} \sum_{k=1}^n A(\beta, Z_k)\frac{\partial}{\partial\beta}\mu(\beta, Z_k).
    \end{split}
    \end{equation}
    Condition 3. together with the dominated convergence theorem implies
    \begin{enumerate}
        \item[(i)] $M(\beta)$ is continuous,
        \item[(ii)] $E(\sup_{\beta\in G}||\frac{\partial}{\partial\beta}(A(\beta, Z_1))(\hat{\theta}^{\ast}_{n,1} -\mu(\beta, Z_1))||)<\infty,$
        \item[(iii)] $E(\sup_{\beta\in G}||A(\beta, Z_1)\frac{\partial}{\partial\beta}\mu(\beta, Z_1)||)<\infty.$
    \end{enumerate}
    By the triangle inequality we have
    \begin{equation*}
        \begin{split}
            &\sup_{\beta\in G}||\frac{\partial}{\partial \beta}(\frac{1}{n}U_n^{\ast}(\beta))-M(\beta_0)||\\
            \leq& \sup_{\beta\in G}||\frac{\partial}{\partial \beta}(\frac{1}{n}U_n^{\ast}(\beta))-M(\beta)||
            +\sup_{\beta\in G}||M(\beta)-M(\beta_0)||\\
            \leq&\sup_{\beta\in G}||\frac{1}{n}\sum_{k=1}^n \frac{\partial}{\partial\beta}(A(\beta, Z_k))(\hat{\theta}^{\ast}_{n,k} -\mu(\beta, Z_k))||\\
            &+ \sup_{\beta\in G}||- \frac{1}{n} \sum_{k=1}^n A(\beta, Z_k)\frac{\partial}{\partial\beta}\mu(\beta, Z_k)-M(\beta)|| +\sup_{\beta\in G}||M(\beta)-M(\beta_0)||.
        \end{split}
    \end{equation*}
    The first term converges a.s. to 0 by Theorem \ref{Uniform LLN U and V} (a), because it follows from the Equations (3.30) and (3.31) in \citet{overgaard2017}, that $$\frac{1}{n}\sum_{k=1}^n \frac{\partial}{\partial\beta}(A(\beta, Z_k))(\hat{\theta}^{\ast}_{n,k} -\mu(\beta, Z_k))$$
    is a mean zero U-Statistic of degree 2 with kernel
    \begin{equation}\label{kernelderivativematrix}
        \begin{split}
            h_{\beta}(z_1,x_1,z_2,x_2)=&\frac{1}{2}\Big(\sum_{i=1}^2(\frac{\partial}{\partial\beta}A(\beta,z_i))(\dot\phi(x_i)-E(\dot\phi(X)|Z=z_i))\\
        &+(\frac{\partial}{\partial\beta}A(\beta,z_1)+\frac{\partial}{\partial\beta}A(\beta,z_2))\ddot\phi(x_1,x_2)\Big).
        \end{split}
    \end{equation}
    As means are U-Statistics of degree 1, the second term converges a.s. to 0 as well. The last term can be made arbitrary small by decreasing the radius of $G$, because continuous functions are uniformly continuous on compact sets.
    
    It remains to show, that we can approximate $\frac{\partial}{\partial \beta}(\frac{1}{n}U_n(\beta))$ by $\frac{\partial}{\partial \beta}(\frac{1}{n}U_n^{\ast}(\beta))$ uniformly well. 
    Define $\Tilde{M}(\beta)=E(\frac{\partial}{\partial\beta}A(\beta, Z_1))$, which is continuous due to condition 3. by the dominated convergence theorem. This implies $$E(\sup_{\beta\in G}||\frac{\partial}{\partial\beta}A(\beta, Z_1)||)<\infty.$$ Therefore we have
    \begin{equation*}
        \begin{split}
            &\sup_{\beta\in G}||\frac{\partial}{\partial \beta}(\frac{1}{n}U_n^{\ast}(\beta))-\frac{\partial}{\partial \beta}(\frac{1}{n}U_n(\beta))||\\
            =&\sup_{\beta\in G}||\frac{1}{n}\sum_{k=1}^n\frac{\partial}{\partial\beta}(A(\beta, Z_k))(\hat{\theta}^{\ast}_{n,k} -\hat{\theta}_{n,k})||\\
            \leq& \max_{k\leq n}|R_{n,k}|\frac{1}{n}\sum_{k=1}^n\sup_{\beta\in G}||\frac{\partial}{\partial\beta}A(\beta, Z_k)||=o_p(1)O_p(1)
        \end{split}
    \end{equation*}
    by the law of large numbers, which completes the proof.
\end{proof}

\begin{remark}
   This also proves Lemma \ref{M estimator}.
\end{remark}

\begin{proof}[Proof of Theorem \ref{HW convergence}]

We consider a fixed but arbitrary element of the estimator. For $i,j\in \{1,\dots,q\}$ by inserting the expansion of the pseudo-observations we have:

\begin{equation}\label{HW expansion}
    \begin{split}
        \Hat{\Sigma}^{HW}_{n,ij}=& \frac{1}{n}\sum_{k=1}^n A(\Hat{\beta}_n, Z_k)_iA(\Hat{\beta}_n, Z_k)_j(\hat{\theta}_{n,k} -\mu(\Hat{\beta}_n, Z_k))^2\\
        =& \frac{1}{n}\sum_{k=1}^n A(\Hat{\beta}_n, Z_k)_iA(\Hat{\beta}_n, Z_k)_j(\phi(F) + \dot \phi(X_k)\\
        &+\frac{1}{n-1}\sum_{k\ne l}\ddot\phi(X_k, X_l) + R_{n,k}-\mu(\Hat{\beta}_n, Z_k))^2\\
        =& \frac{1}{n}\sum_{k=1}^n A(\Hat{\beta}_n, Z_k)_iA(\Hat{\beta}_n, Z_k)_j(\phi(F) + \dot \phi(X_k)-\mu(\Hat{\beta}_n, Z_k))^2 \\
        &+ \frac{1}{n}\sum_{k=1}^n A(\Hat{\beta}_n, Z_k)_iA(\Hat{\beta}_n, Z_k)_j(\frac{1}{n-1}\sum_{k\ne l}\ddot\phi(X_k, X_l))^2\\
        &+\frac{1}{n}\sum_{k=1}^n A(\Hat{\beta}_n, Z_k)_iA(\Hat{\beta}_n, Z_k)_jR_{n,k}^2\\
        &+ \frac{2}{\binom{n}{2}}\sum_{k=1}^n\sum_{k\ne l} A(\Hat{\beta}_n, Z_k)_iA(\Hat{\beta}_n, Z_k)_j(\phi(F) + \dot \phi(X_k)-\mu(\Hat{\beta}_n, Z_k))\ddot\phi(X_k, X_l)\\
        &+ \frac{2}{n}\sum_{k=1}^n A(\Hat{\beta}_n, Z_k)_iA(\Hat{\beta}_n, Z_k)_j(\phi(F) + \dot \phi(X_k)-\mu(\Hat{\beta}_n, Z_k))R_{n,k}\\
        &+ \frac{2}{\binom{n}{2}}\sum_{k=1}^n\sum_{k\ne l} A(\Hat{\beta}_n, Z_k)_iA(\Hat{\beta}_n, Z_k)_jR_{n,k}\ddot\phi(X_k, X_l)\\
    \end{split}
\end{equation}
If we would ignore the fact this expression still depends on $\Hat{\beta}_n$, the first term is an i.i.d. representation of the estimator. It therefore remains to show that the remaining terms converge to 0 in probability and that the use of the estimator $\Hat{\beta}_n$ instead of $\beta_0$ leads to the same limit. Due to the Cauchy-Schwarz inequality, it is sufficient to show that the second and third term converge in probability to 0, as long as the first is stochastically bounded. Define $G=\overline{B}_{\varepsilon}(\beta_0)$ as in the proof of Theorem \ref{overgaard main variant}. Because $\Hat{\beta}_n$ is consistent, the probability of the event $\{\Hat{\beta}_n \in G\}$ tends to 1. It is, as in the proof of Theorem \ref{overgaard main variant}, no restriction to assume, that the dominated integrability conditions holds on this set as well.

For the second third term in \eqref{HW expansion} we have for large $n$ with high probability
\begin{equation*}
    \begin{split}
        \frac{1}{n}\sum_{k=1}^n A(\Hat{\beta}_n, Z_k)_iA(\Hat{\beta}_n, Z_k)_jR_{n,k}^2&\leq\frac{1}{n}\sum_{k=1}^n \sup_{\beta\in G} |A(\beta, Z_k)_iA(\beta, Z_k)_jR_{n,k}^2|\\
        &\leq \max_{k\leq n} R_{n,k}^2\frac{1}{n}\sum_{k=1}^n \sup_{\beta\in G}|A(\beta_0, Z_k)_iA(\beta_0, Z_k)_j|\\
        &=o_p(1)O_p(1)
    \end{split}
\end{equation*}
by local dominated square integrability of $A(\beta,Z)$ and the law of large numbers. To analyze $$J_1=\frac{1}{n}\sum_{k=1}^n A(\Hat{\beta}_n, Z_k)_iA(\Hat{\beta}_n, Z_k)_j(\frac{1}{n-1}\sum_{k\ne l}\ddot\phi(X_k, X_l))^2$$ we define
\begin{equation*}
    \begin{split}
        \psi_{\beta}(Z_1,X_1,Z_2,X_2,Z_3,X_3)=&\frac{1}{3}(A(\beta, Z_1)_iA(\beta, Z_1)_j+A(\beta, Z_2)_iA(\beta, Z_2)_j\\
        &+A(\beta, Z_3)_iA(\beta, Z_3)_j)\ddot\phi(X_1, X_2)\ddot\phi(X_1, X_3),
    \end{split}
\end{equation*}
and its symmetrization
\begin{equation*}
    \begin{split}
       \Tilde{\psi}_{\beta}(Z_1,X_1,Z_2,X_2,Z_3,X_3)=&\frac{1}{3}(\psi_{\beta}(Z_1,X_1,Z_2,X_2,Z_3,X_3)\\
       &+\psi_{\beta}(Z_2,X_2,Z_1,X_1,Z_3,X_3)\\
       &+\psi_{\beta}(Z_3,X_3,Z_2,X_2,Z_1,X_1)).
    \end{split}
\end{equation*}
 We can represent $J_1$ in terms of these functions as follows
\begin{equation*}
    \begin{split}
        J_1=&\frac{1}{n}\sum_{k=1}^n A(\Hat{\beta}_n, Z_k)_iA(\Hat{\beta}_n, Z_k)_j(\frac{1}{n-1}\sum_{k\ne l}\ddot\phi(X_k, X_l))^2\\
        =& 
        \frac{1}{n(n-1)^2}\sum_{k=1}^n\mathop{\underset{l\ne r}{\sum_{k\ne l}\sum_{k\ne r}}}A(\Hat{\beta}_n, Z_k)_iA(\Hat{\beta}_n, Z_k)_j\ddot\phi(X_k, X_l)\ddot\phi(X_k, X_r)\\
        &+\frac{1}{n(n-1)^2}\sum_{k=1}^n\sum_{k\ne l}A(\Hat{\beta}_n, Z_k)_iA(\Hat{\beta}_n, Z_k)_j\ddot\phi(X_k, X_l)^2\\
        =& \frac{n-2}{n-1}\frac{1}{\binom{n}{3}}\sum_{k=1}^n\sum\limits_{\substack{l=1 \\ k\ne l \ne r }}^n\sum_{r=1}^n\psi_{\Hat{\beta}_n}(Z_k,X_k,Z_l,X_l,Z_r,X_r)\\
        &+\frac{2}{n(n-1)^2}\sum_{k=1}^n\sum_{k\ne l}A(\Hat{\beta}_n, Z_k)_iA(\Hat{\beta}_n, Z_k)_j\ddot\phi(X_k, X_l)^2\\
        =& \frac{n-2}{n-1}\frac{1}{\binom{n}{3}}\sum\sum\limits_{1\leq k < l < r \leq n}\sum2\Tilde{\psi}_{\Hat{\beta}_n}(Z_k,X_k,Z_l,X_l,Z_r,X_r)\\
        &+\frac{2}{n(n-1)^2}\sum_{k=1}^n\sum_{k\ne l}A(\Hat{\beta}_n, Z_k)_iA(\Hat{\beta}_n, Z_k)_j\ddot\phi(X_k, X_l)^2\\
        =& \frac{n-2}{n-1} 2U_n^{(3)}(\Tilde{\psi}_{\Hat{\beta}_n})\\
        &+\frac{2}{n(n-1)^2}\sum_{k=1}^n\sum_{k\ne l}A(\Hat{\beta}_n, Z_k)_iA(\Hat{\beta}_n, Z_k)_j\ddot\phi(X_k, X_l)^2.\\
    \end{split}
\end{equation*}

Here $U_n^{(3)}(\Tilde{\psi}_{\Hat{\beta}_n})$ is a U-statistic of degree 3 with kernel $\Tilde{\psi}_{\Hat{\beta}_n}(Z_1,X_1,Z_2,X_2,Z_3,X_3)$. Because $$E(\sup_{\beta\in G}|\Tilde{\psi}_{\beta}(Z_1,X_1,Z_2,X_2,Z_3,X_3)|)<\infty$$ by our regularity assumptions, we have as $n \to \infty$
\[\sup_{\beta\in G}|U_n^{(3)}(\Tilde{\psi}_{\beta})-E(\Tilde{\psi}_{\beta_0}(Z_1,X_1,Z_2,X_2,Z_3,X_3))|\xrightarrow{\text{a.s.}}0\]
by Theorem \ref{Uniform LLN U and V} (a) and a uniform continuity argument as in the proof of Theorem \ref{overgaard main variant}.
Additionally we have for a constant $K < \infty$
\begin{equation*}
    \begin{split}
        &\left|\frac{2}{n(n-1)^2}\sum_{k=1}^n\sum_{k\ne l}A(\Hat{\beta}_n, Z_k)_iA(\Hat{\beta}_n, Z_k)_j\ddot\phi(X_k, X_l)^2\right|\\
        \leq& K\frac{2}{n(n-1)}\sum_{k=1}^n\sup_{\beta\in G}|A(\beta, Z_k)_iA(\beta, Z_k)_j|\\
        =& O_p(n^{-1})
    \end{split}
\end{equation*}
by the law of large numbers. It therefore remains to show, that $U_n^{(3)}(\Tilde{\psi}_{\beta_0})$ is centered. First, note that $E(\Tilde{\psi}_{\beta_0}(Z_1,X_1,Z_2,X_2,Z_3,X_3)) = E(\psi_{\beta_0}(Z_1,X_1,Z_2,X_2,Z_3,X_3))$. Due to the independence of $(X_1,Z_1), \dots, (X_n,Z_n)$ it holds
\begin{equation*}
    \begin{split}&E(\psi_{\beta_0}(Z_1,X_1,Z_2,X_2,Z_3,X_3))\\
    &=E(E(\psi_{\beta_0}(Z_1,X_1,Z_2,X_2,Z_3,X_3)|X_1))\\
        &=\frac{1}{3}\Big(E(E(A(\beta_0, Z_1)_iA(\beta_0, Z_1)_j\ddot\phi(X_1, X_2)\ddot\phi(X_1, X_3)| X_1))\\
        &\phantom{=}+E(E(A(\beta_0, Z_2)_iA(\beta_0, Z_2)_j\ddot\phi(X_1, X_2)\ddot\phi(X_1, X_3)| X_1))\\
        &\phantom{=}+E(E(A(\beta_0, Z_3)_iA(\beta_0, Z_3)_j\ddot\phi(X_1, X_2)\ddot\phi(X_1, X_3)| X_1))\Big)\\
        &=\frac{1}{3}\Big(E(E(A(\beta_0, Z_1)_iA(\beta_0, Z_1)_j|X_1))E(\ddot\phi(X_1, X_2))E(\ddot\phi(X_1, X_3))\\
        &\phantom{=}+E(A(\beta_0, Z_2)_iA(\beta_0, Z_2)_j\ddot\phi(X_1, X_2))E(\ddot\phi(X_1, X_3))\\
        &\phantom{=}+E(A(\beta_0, Z_3)_iA(\beta_0, Z_3)_j\ddot\phi(X_1, X_3))E(\ddot\phi(X_1, X_2))\Big)\\
        &=0
    \end{split}
\end{equation*}
because $\ddot\phi(.,.)$ is centered (see Equation 3.24 in \citet{overgaard2017}). Combining these results, we have
\begin{equation*}
    \Hat{\Sigma}^{\text{HW}}_{n,ij} = \frac{1}{n}\sum_{k=1}^n A(\Hat{\beta}_n, Z_k)_iA(\Hat{\beta}_n, Z_k)_j(\phi(F) + \dot \phi(X_k)-\mu(\Hat{\beta}_n, Z_k))^2 +o_p(1),
\end{equation*}
and therefore $\Hat{\Sigma}^{\text{HW}}_{n,ij} \xrightarrow{p}\text{Cov}\left(A(\beta_0, Z_1)(\phi(F)+\dot \phi(X_1)-\mu(\beta, Z_1))\right)_{ij}$
with another application of Theorem \ref{Uniform LLN U and V} (a) which completes the proof.
\end{proof}

\begin{proof}[Proof of Theorem \ref{Consistency corrected variance est}]
     As in the proof of Theorem \ref{HW convergence}, we consider a fixed but arbitrary element of the estimator. Expanding the formula leads to
    \begin{equation*}\label{PV_estimator_elements}
        \begin{split}
            \Hat{\Sigma}^{PV}_{n,ij}=&\frac{1}{n}\sum_{k=1}^n\Bigg(A(\Hat{\beta}_n,Z_k)_iA(\Hat{\beta}_n,Z_k)_j(\phi(F_n))^2\\
            &+A(\Hat{\beta}_n,Z_k)_iA(\Hat{\beta}_n,Z_k)_j(\phi'_{F_n}(\delta_{X_k}-F_n))^2\\
            &+A(\Hat{\beta}_n,Z_k)_iA(\Hat{\beta}_n,Z_k)_j(\mu(\Hat{\beta}_n,Z_k))^2\\
            &-2A(\Hat{\beta}_n,Z_k)_iA(\Hat{\beta}_n,Z_k)_j\phi(F_n)\mu(\Hat{\beta}_n,Z_k)\\
            &+2A(\Hat{\beta}_n,Z_k)_iA(\Hat{\beta}_n,Z_k)_j\phi(F_n)\phi'_{F_n}(\delta_{X_k}-F_n)\\
            &-2A(\Hat{\beta}_n,Z_k)_iA(\Hat{\beta}_n,Z_k)_j\mu(\Hat{\beta}_n,Z_k)\phi'_{F_n}(\delta_{X_k}-F_n)\\
            &+\Hat{h}_1(X_k)_i\Hat{h}_1(X_k)_j\\
            &+2A(\Hat{\beta}_n,Z_k)_i\Hat{h}_1(X_k)_j\phi'_{F_n}(\delta_{X_k}-F_n)\\
            &+2A(\Hat{\beta}_n,Z_k)_i\Hat{h}_1(X_k)_j\phi(F_n)\\
            &-2A(\Hat{\beta}_n,Z_k)_i\Hat{h}_1(X_k)_j\mu(\Hat{\beta}_n,Z_k)\Bigg)\\
            =&\Hat{\Sigma}^{(1)}_{n,ij}(\Hat{\beta}_n, F_n)+\Hat{\Sigma}^{(2)}_{n,ij}(\Hat{\beta}_n, F_n)+2\Hat{\Sigma}^{(3)}_{n,ij}(\Hat{\beta}_n, F_n),
        \end{split}
    \end{equation*}
    with
        \begin{equation*}
        \begin{split}
            \Hat{\Sigma}^{(1)}_{n,ij}(\Hat{\beta}_n, F_n)=\frac{1}{n}\sum_{k=1}^n\Bigg(&A(\Hat{\beta}_n,Z_k)_iA(\Hat{\beta}_n,Z_k)_j(\phi(F_n))^2\\
            &+A(\Hat{\beta}_n,Z_k)_iA(\Hat{\beta}_n,Z_k)_j(\phi_{F_n}'(\delta_{X_k}-F_n))^2\\
            &+A(\Hat{\beta}_n,Z_k)_iA(\Hat{\beta}_n,Z_k)_j(\mu(\Hat{\beta}_n,Z_k))^2\\
            &-2A(\Hat{\beta}_n,Z_k)_iA(\Hat{\beta}_n,Z_k)_j\phi(F_n)\mu(\Hat{\beta}_n,Z_k)\\
            &+2A(\Hat{\beta}_n,Z_k)_iA(\Hat{\beta}_n,Z_k)_j\phi(F_n)\phi'_{F_n}(\delta_{X_k}-F_n)\\
            &-2A(\Hat{\beta}_n,Z_k)_iA(\Hat{\beta}_n,Z_k)_j\mu(\Hat{\beta}_n,Z_k)\phi'_{F_n}(\delta_{X_k}-F_n)\Bigg),
        \end{split}
    \end{equation*}

        \begin{equation*}
        \begin{split}
            \Hat{\Sigma}^{(2)}_{n,ij}(\Hat{\beta}_n, F_n)=\frac{1}{n^3}\sum_{k,l,r=1}^nA(\Hat{\beta}_n,Z_l)_iA(\Hat{\beta}_n,Z_r)_j\phi''_{F_n}(\delta_{X_l}-F_n,\delta_{X_k}-F_n)\phi''_{F_n}(\delta_{X_r}-F_n,\delta_{X_k}-F_n),
        \end{split}
    \end{equation*}

        \begin{equation*}
        \begin{split}
            \Hat{\Sigma}^{(3)}_{n,ij}(\Hat{\beta}_n, F_n)=\frac{1}{n^2}\sum_{k,l=1}^n\Bigg(
            &A(\Hat{\beta}_n,Z_k)_iA(\Hat{\beta}_n,Z_l)_j\phi''_{F_n}(\delta_{X_l}-F_n,\delta_{X_k}-F_n)\phi'_{F_n}(\delta_{X_k}-F_n)\\
            &+A(\Hat{\beta}_n,Z_k)_iA(\Hat{\beta}_n,Z_l)_j\phi''_{F_n}(\delta_{X_l}-F_n,\delta_{X_k}-F_n)\phi(F_n)\\
            &-A(\Hat{\beta}_n,Z_k)_iA(\Hat{\beta}_n,Z_l)_j\phi''_{F_n}(\delta_{X_l}-F_n,\delta_{X_k}-F_n)\mu(\Hat{\beta}_n,Z_k)\Bigg).
        \end{split}
    \end{equation*}
     Here we made the dependence of the estimator on $\Hat{\beta}_n$ and $F_n$ explicit. To handle the dependence of estimator on $F_n$, we take a closer look at the behavior of the derivatives of the functional $\phi.$ Considering $\phi'',$
    we have due to linearity
    \begin{equation}\label{second derivative difference}
    \begin{split}
        &\phi_{F_n}''(\delta_{X_k}-F_n,\delta_{X_j}-F_n)-\phi_{F}''(\delta_{X_k}-F,\delta_{X_j}-F)\\
        =&\phi_{F_n}''(\delta_{X_k}-F_n,\delta_{X_j}-F_n)+\phi_{F_n}''(\delta_{X_k}-F,\delta_{X_j}-F)\\&-\phi_{F_n}''(\delta_{X_k}-F,\delta_{X_j}-F)-\phi_{F}''(\delta_{X_k}-F,\delta_{X_j}-F)\\
        &+\phi_{F}''(F_n-F,F_n-F)-\phi_{F}''(F_n-F,F_n-F)\\
        =&(\phi_{F_n}''-\phi_{F}'')(F_n-F,F_n-F)+(\phi_{F_n}''-\phi_{F}'')(\delta_{X_k}-F,\delta_{X_j}-F)\\&+\phi_{F}''(F_n-F,F_n-F).
    \end{split}
    \end{equation}
    By the assumption of local Lipschitz continuity which also implies that $\phi''$ is locally bounded, it holds for large $n$ with high probability
    \begin{equation*}
        \begin{split}
            &||\phi_{F_n}''(\delta_{X_k}-F_n,\delta_{X_j}-F_n)-\phi_{F}''(\delta_{X_k}-F,\delta_{X_j}-F)||\\
            \leq&K||F_n-F||^3+K||F_n-F||(\max_{l}||\delta_{X_l}||+||F||)^2+K||F_n-F||^2
        \end{split}
    \end{equation*}
    for a constant $K<\infty.$ Similarly, it follows that
        \begin{equation*}
        \begin{split}
            &||\phi_{F_n}'(\delta_{X_k}-F_n)-\phi_{F}'(\delta_{X_k}-F)||\\
            \leq&K||F_n-F||^2+K||F_n-F||(\max_{l}||\delta_{X_l}||+||F||)+K||F_n-F||.
        \end{split}
    \end{equation*}
    We only consider $\Hat{\Sigma}^{(2)}_{n,ij}(\Hat{\beta}_n, F_n)$ in detail, the other two terms can be handled similarly. As in the previous proofs, let $G=\overline{B}_{\varepsilon}(\beta_0).$ Additionally define 
    \[\Sigma^{(2)}_{ij}(\beta, F)=\text{Cov}(A(\beta,Z_2)\ddot\phi(X_1, X_2)|X_1))_{ij}.\]
    Because $\Hat{\beta}_n$ is consistent for $\beta_0$, the probability of the event $\Hat{\beta}_n\in G$ tends to 1. Similarly, $F_n$ is close to $F$ with a probability tending to 1. Thus, on the intersection of both events it holds
    \begin{equation}\label{Sigma2 Hat decomposition}
    \begin{split}
        &|\Hat{\Sigma}^{(2)}_{n,ij}(\Hat{\beta}_n, F_n)-\Sigma^{(2)}_{ij}(\beta_0, F)|\\
        =& |\Hat{\Sigma}^{(2)}_{n,ij}(\Hat{\beta}_n, F_n)- \Hat{\Sigma}^{(2)}_{n,ij}(\Hat{\beta}_n, F) +\Hat{\Sigma}^{(2)}_{n,ij}(\Hat{\beta}_n, F) -\Sigma^{(2)}_{ij}(\beta_0, F)|\\
        \leq&|\Hat{\Sigma}^{(2)}_{n,ij}(\Hat{\beta}_n, F_n)- \Hat{\Sigma}^{(2)}_{n,ij}(\Hat{\beta}_n, F)|\\
        &+\sup_{\beta \in G}|\Hat{\Sigma}^{(2)}_{n,ij}(\beta, F)-\Sigma^{(2)}_{ij}(\beta, F)|\\
        &+\sup_{\beta \in G}|\Sigma^{(2)}_{ij}(\beta, F)-\Sigma^{(2)}_{ij}(\beta_0, F)|.
    \end{split}
    \end{equation}
    Due to \eqref{second derivative difference} we have for the first term
    \begin{equation*}
        \begin{split}
            &|\Hat{\Sigma}^{(2)}_{n,ij}(\Hat{\beta}_n, F_n)- \Hat{\Sigma}^{(2)}_{n,ij}(\Hat{\beta}_n, F)|\\
            =&\Big|\frac{1}{n^3}\sum_{k,l,r=1}^nA(\Hat{\beta}_n,Z_l)_iA(\Hat{\beta}_n,Z_r)_j\\
            &\times \left(\phi''_{F_n}(\delta_{X_l}-F_n,\delta_{X_k}-F_n)\phi''_{F_n}(\delta_{X_r}-F_n,\delta_{X_k}-F_n)-\ddot\phi(X_l, X_k)\ddot\phi(X_r, X_k)\right)\Big|\\
            \leq&\Big|\frac{1}{n^3}\sum_{k,l,r=1}^nA(\Hat{\beta}_n,Z_l)_iA(\Hat{\beta}_n,Z_r)_j\\
            &\times \left(\phi''_{F_n}(\delta_{X_l}-F_n,\delta_{X_k}-F_n)(\phi''_{F_n}(\delta_{X_r}-F_n,\delta_{X_k}-F_n)-\ddot\phi(X_r, X_k))\right)\Big|\\
            &+\Big|\frac{1}{n^3}\sum_{k,l,r=1}^nA(\Hat{\beta}_n,Z_l)_iA(\Hat{\beta}_n,Z_r)_j\times \left(\ddot\phi(X_r, X_k)(\phi''_{F_n}(\delta_{X_l}-F_n,\delta_{X_k}-F_n)-\ddot\phi(X_l, X_k))\right)\Big|\\
            \leq&2K\left(K||F_n-F||^3+K||F_n-F||(\max_{k\leq n}||\delta_{X_k}||+||F||)^2+K||F_n-F||^2\right)\\
            &\times \frac{1}{n^2}\sum_{l,r=1}^n\sup_{\beta\in G}|A(\beta,Z_l)_iA(\beta,Z_r)_j|\\
            =&o_p(1)O_p(1)
        \end{split}
    \end{equation*}
    by the law of large numbers for V-statistics, which is applicable because $A(\beta, Z)$ is locally dominated square integrable.
    With a similar calculation as in the proof of Lemma \ref{HW convergence}, we can represent $\Hat{\Sigma}^{(2)}_{ij}(\beta, F)$ with the function 
    \begin{equation*}
        \begin{split}
           \Psi_{\beta}(Z_1,X_1,Z_2,X_2,Z_3,X_3)= \frac{1}{3}\Big(&A(\beta_0, Z_1)_iA(\beta_0, Z_2)_j\ddot\phi(X_1, X_3)\ddot\phi(X_2, X_3)\\
           &+A(\beta_0, Z_2)_iA(\beta_0, Z_3)_j\ddot\phi(X_1, X_2)\ddot\phi(X_1, X_3)\\
        &+A(\beta_0, Z_1)_iA(\beta_0, Z_3)_j\ddot\phi(X_1, X_2)\ddot\phi(X_2, X_3)\Big)
        \end{split}
    \end{equation*}
    as
    \begin{equation*}
        \Hat{\Sigma}^{(2)}_{n,ij}(\beta, F)= \frac{1}{n^3}\sum_{k,l,r=1}^n\Psi_{\beta}(Z_k,X_k,Z_l,X_l,Z_r,X_r).
    \end{equation*}
    This is a V-Statistic of order 3 with kernel $\Psi_{\beta}.$ Due to the dominated integrability  assumptions on $A(\beta,Z)$, we can apply the uniform law of large numbers for V-Statistics (Lemma \ref{Uniform LLN U and V} (b)), which implies
    \begin{equation*}
        \sup_{\beta\in G}|\Hat{\Sigma}^{(2)}_{n,ij}(\beta, F)-E(\Psi_{\beta}(Z_1,X_1,Z_2,X_2,Z_3,X_3))|\xrightarrow{\text{a.s.}}0.
    \end{equation*}
By the tower property for conditional expectations, we have
\begin{equation*}
\begin{split}
    &E(\Psi_{\beta}(Z_1,X_1,Z_2,X_2,Z_3,X_3))\\=&E\Big(E(A(\beta,Z_2)_i\ddot\phi(X_1, X_2)|X_1)E(A(\beta,Z_3)_j\ddot\phi(X_1, X_3)|X_1)\Big)
\end{split}
\end{equation*}
as well as
\begin{equation*}
    \begin{split}
       &E(E(A(\beta,Z_2)_i\ddot\phi(X_1, X_2)|X_1))\\
       =&E(E(A(\beta,Z_2)_i\ddot\phi(X_1, X_2)|Z_2))\\
       =&E(A(\beta,Z_2)_i)E(E(\ddot\phi(X_1, X_2)|Z_2))\\
       =&E(A(\beta,Z_2)_i)E(E(\ddot\phi(X_1, X_2)|X_1))\\
       =&0.
    \end{split}
\end{equation*}
Therefore \begin{equation*}
    E(\Psi_\beta(Z_1,X_1,Z_2,X_2,Z_3,X_3))=\text{Cov}(A(\beta,Z_2)\ddot\phi(X_1, X_2)|X_1))_{ij}.
\end{equation*}
As in the previous proofs, the last term of \eqref{Sigma2 Hat decomposition} can be made arbitrarily small by decreasing the radius of $G.$ Summarizing, we have $$\Hat{\Sigma}^{(2)}_{n,ij}(\Hat{\beta}_n, F_n)\xrightarrow{p}\text{Cov}(A(\beta_0,Z_2)\ddot\phi(X_1, X_2)|X_1))_{ij}.$$

Similarly, it follows that $\Hat{\Sigma}^{(3)}_{ij}(\beta, F)$ is a sum of 3 V-Statistics, each of order~2, which implies the convergence
\begin{equation*}
    \begin{split}
        \Hat{\Sigma}^{(3)}_{ij}(\Hat{\beta}_n, F_n)\xrightarrow{p}E\Bigg(
            &A(\beta_0,Z_1)_i\dot\phi(X_1)E(A(\beta_0,Z_2)_j\ddot\phi(X_2,X_1)|X_1)\\
            &+A(\beta_0,Z_1)_i\phi(F)E(A(\beta_0,Z_2)_j\ddot\phi(X_2,X_1)|X_1)\\
            &-A(\beta_0,Z_1)_i\mu(\beta_0,Z_1)E(A(\beta_0,Z_2)_j\ddot\phi(X_2,X_1)|X_1)\Bigg)\\
            =&E\Big(A(\beta_0,Z_1)_i(\phi(F)+\dot\phi(X_1)-\mu(\beta_0,Z_1))\\&\times E(A(\beta_0,Z_2)_j\ddot\phi(X_2,X_1)|X_1)\Big)\\
            =&\text{Cov}\Big(A(\beta_0,Z_1)(\phi(F)+\dot\phi(X_1)-\mu(\beta_0,Z_1)),\\&E(A(\beta_0,Z_2)\ddot\phi(X_2,X_1)|X_1)\Big)_{ij}.
    \end{split}
\end{equation*}
Additionally, $\Hat{\Sigma}^{(1)}_{n,ij}(\beta, F)$ is a simple average of i.i.d.\ terms. Therefore,
     \begin{equation*}
         \begin{split}
             \Hat{\Sigma}^{(1)}_{n,ij}(\Hat{\beta}_n, F_n)\xrightarrow{p}E\Bigg(&A(\beta_0,Z_1)_iA(\beta_0,Z_1)_j(\phi(F))^2\\
             &+A(\beta_0,Z_1)_iA(\beta_0,Z_1)_j(\dot\phi(X_1))^2\\
             &+A(\beta_0,Z_1)_iA(\beta_0,Z_1)_j(\mu(\beta_0,Z_1))^2\\
             &-2A(\beta_0,Z_1)_iA(\beta_0,Z_1)_j\phi(F)\mu(\beta_0,Z_1)\\
             &+2A(\beta_0,Z_1)_iA(\beta_0,Z_1)_j\phi(F)\dot\phi(X_1)\\
             &-2A(\beta_0,Z_1)_iA(\beta_0,Z_1)_j\mu(\beta_0,Z_1)\dot\phi(X)\Bigg)\\
             =&E\Big(A(\beta_0,Z_1)_i(\phi(F)+\dot\phi(X_1)-\mu(\beta_0,Z_1))\\
             &\times A(\beta_0,Z_1)_j(\phi(F)+\dot\phi(X_1)-\mu(\beta_0,Z_1))\Big)\\
             =&\text{Cov}\Big(A(\beta_0,Z_1)(\phi(F)+\dot\phi(X_1)-\mu(\beta_0,Z_1))\Big)_{ij}.
         \end{split}
     \end{equation*}

Combining the previous results, we get
\begin{equation*}
    \begin{split}
        \Hat{\Sigma}^{PV}_{n,ij}\xrightarrow{p} \text{Cov}&\Big(A(\beta_0,Z_1)(\phi(F)+\dot\phi(X_1)-\mu(\beta_0,Z_1))\Big)_{ij}\\
        &+\text{Cov}(A(\beta_0,Z_2)\ddot\phi(X_1, X_2)|X_1))_{ij}\\
        &+2\text{Cov}\Big(A(\beta_0,Z_1)(\phi(F)+\dot\phi(X_1)-\mu(\beta_0,Z_1)),\\&\quad E(A(\beta_0,Z_2)\ddot\phi(X_2,X_1)|X_1)\Big)_{ij}\\
        =&\text{Cov}\Big(A(\beta_0,Z_1)\big(\phi(F)+\dot\phi(X_1)-\mu(\beta_0,Z_1)\big)\\&+ E(A(\beta_0,Z_2)\ddot\phi(X_2,X_1)|X_1)\Big)_{ij}\\
        =&\Sigma_{ij}
    \end{split}
\end{equation*}
which completes the proof.
\end{proof}

\begin{proof}[Proof of Theorem \ref{asymptotic Wald test}]
    It follows from Theorem \ref{overgaard main variant} that $\sqrt{n}(\Hat{\beta}_n-\beta_0)$ converges in distribution to $N_q(0,M^{-1}\Sigma (M^{-1})^T).$ Additionally, it follows from Lemma \ref{M estimator} and Theorem \ref{Consistency corrected variance est} that $\Hat{M}_n$ and $\Hat{\Sigma}_n^{PV}$ converge in probability to $M$ and $\Sigma$ respectively. Moreover, we can w.l.o.g. assume that $\Hat{M}_n$ and $\Hat{\Sigma}_n^{PV}$ converge almost surely, otherwise pass to an almost surely convergent subsequence. Because $M^{-1}\Sigma (M^{-1})^T$ is symmetric and positive definite, $$\text{rank}(CM^{-1}\Sigma (M^{-1})^TC^T)=\text{rank}(C(M^{-1}\Sigma (M^{-1})^T)^{1/2})= \text{rank}(C).$$ Similarly $\text{rank}(C\Hat{M}^{-1}_n\Hat{\Sigma}_n^{PV}(\Hat{M}^{-1}_n)^TC^T)=\text{rank}(C)$ almost surely for sufficiently large $n.$ Therefore there is (almost surely) no rank jump in $(C\Hat{M}_n^{-1}\Hat{\Sigma}_n^{PV}(\Hat{M}_n^{-1})^TC^T)$, which implies $$C\Hat{M}_n^{-1}\Hat{\Sigma}_n^{PV}(\Hat{M}_n^{-1})^TC^T)^{+} \xrightarrow{\text{a.s.}}(CM^{-1}\Sigma (M^{-1})^TC^T)^+$$ by Corollary 1.8 of \citet{koliha2001continuity} and the continuous mapping theorem. Therefore (a) follows from Theorem 7.3 in \citet{rao1972generalized}, together with Slutzky's lemma.
    
    Part (b) can be proven similarly to Theorem 2 in \citet{ditzhaus2021qanova}. Due to the consistency of $\Hat{\beta}_n,\Hat{M}_n$ and $\Hat{\Sigma}_n^{PV}$
\[(C\hat{\beta}_n-b)^T(C\Hat{M}_n^{-1}\Hat{\Sigma}_n^{PV}(\Hat{M}_n^{-1})^TC^T)^{+}(C\hat{\beta}_n-b)\xrightarrow{p}(C\beta_0-b)^T(CM^{-1}\Sigma (M^{-1})^TC^T)^{+}(C\beta_0-b)\]
under $H_0$ as well as $H_1.$ Thus it remains to prove that $C\beta_0-b\ne0$ always implies 
\[(C\beta_0-b)^T(CM^{-1}\Sigma (M^{-1})^TC^T)^{+}(C\beta_0-b)>0.\] Define $\Gamma:=M^{-1}\Sigma (M^{-1})^T$ and its symmetric square root $\Gamma^{1/2},$ which is positive definite as well. Therefore there is some $\Tilde{b}$ such that $\Gamma^{1/2}\Tilde{b}=\beta_0.$ Additionally we consider the matrix $C\Gamma^{1/2}(C\Gamma^{1/2})^+$, which is a projection onto the column space of $C\Gamma^{1/2}$ which coincides with the column space of $C.$ This implies $C\Gamma^{1/2}(C\Gamma^{1/2})^+b=b,$ because the system $C\beta_0=b$ is satisfiable. Now recall the well the well known properties of the Moore-Penrose-Inverse $(H^+)^T=(H^T)^+$, $(H^T H)^+=H^+(H^T)^+$ and $HH^+H=H$. First, we have
\begin{equation*}
    \begin{split}
         C\Gamma^{1/2}\left((C\Gamma^{1/2})^+((C\beta_0)-b)\right)&=C\Gamma^{1/2}(C\Gamma^{1/2})^+(C\beta_0)-C\Gamma^{1/2}(C\Gamma^{1/2})^+b\\
         &=C\Gamma^{1/2}\Tilde{b}-b\\
         &=C\beta_0-b\\
         &\ne0.
    \end{split}
\end{equation*}
And thus
\begin{equation*}
    \begin{split}
        (C\beta_0-b)^T(C\Gamma C^T)^{+}(C\beta_0-b)&=(C\beta_0-b)^T(\Gamma^{1/2}C^T)^+(C\Gamma^{1/2})^+(C\beta_0-b)\\
        &=\left((C\Gamma^{1/2})^+(C\beta_0-b)\right)^T\left((C\Gamma^{1/2})^+(C\beta_0-b)\right)\\
        &>0.
    \end{split}
\end{equation*}
\end{proof}

In order to determine the limiting distribution of the bootstrap estimates, we first need a central limit theorem for the bootstrap estimating equation.

\begin{lemma} \label{weak covergence Boot equation}
     Let $(M_{n1}, \dots, M_{nn}) \sim Mult(n, \frac{1}{n}, \dots, \frac{1}{n})$ be a multinomial random vector, independent of the data.
    Assume that  $A(\beta_0, Z)$  has finite second moment. Furthermore assume that there exists some $\delta>0$ such that $E(||A(\beta, Z_1)\mu(\beta, Z_1)||^{2+\delta})< \infty.$ Then, as $n\to\infty$,
    \begin{itemize}
        \item[(a)] \[\frac{1}{\sqrt{n}}\sum_{k=1}^n (M_{nk}-1) A(\beta_0, Z_k)(\hat{\theta}_{n,k} -\mu(\beta_0, Z_k)) \xrightarrow{d} N_q(0, \Tilde{\Sigma})\]
        in probability given the data.
        \item[(b)] \[\frac{1}{\sqrt{n}} U_n^B(\beta_0) \xrightarrow{d} N_q(0, \Sigma + \Tilde{\Sigma}).\]
    \end{itemize}

\end{lemma}

\begin{proof}[Proof of Lemma \ref{weak covergence Boot equation}]
    To prove (a), we apply Theorem 4.1 of \citet{pauly2011martingale}. It follows from Lemma \ref{HW convergence} that as $n\to \infty$
    \begin{equation*}
        \frac{1}{n}\sum_{k=1}^nA(\beta_0, Z_k)A(\beta_0, Z_k)^T(\hat{\theta}_{n,k} -\mu(\beta_0, Z_k))^2 \xrightarrow{p} \Tilde{\Sigma}.
    \end{equation*}
    No dominated integrability conditions are necessary in this case, because in contrast to the Huber-White estimator, the above expression does not depend on $\Hat{\beta}_n.$
    It remains to show that as $n \to \infty$
    \[\max_{k\leq n}||\frac{1}{\sqrt{n}}A(\beta, Z_k)(\hat{\theta}_{n,k} -\mu(\beta, Z_k))||\xrightarrow{p}0.\]
By subadditivity it holds
\begin{equation*}
    \begin{split}
        &P(\max_{k\leq n}||\frac{1}{\sqrt{n}}A(\beta, Z_k)(\hat{\theta}_{n,k} -\mu(\beta, Z_k))||>\varepsilon) \\
        \leq& P(\bigcup_{k=1}^n \{||\frac{1}{\sqrt{n}}A(\beta, Z_k)(\hat{\theta}_{n,k} -\mu(\beta, Z_k))||>\varepsilon\}) \\
        \leq& \sum_{k=1}^n P(||\frac{1}{\sqrt{n}}A(\beta, Z_k)(\hat{\theta}_{n,k} -\mu(\beta, Z_k))||>\varepsilon) \\
        =&\sum_{k=1}^n P(||\frac{1}{\sqrt{n}}A(\beta, Z_k)(\hat{\theta}^{\ast}_{n,k} +R_{n,k} -\mu(\beta, Z_k))||>\varepsilon).\\
    \end{split}
\end{equation*}
Now notice that we have for the term including the remainder
\begin{equation*}
    \begin{split}
        \sum_{k=1}^n \frac{1}{\sqrt{n}}||A(\beta, Z_k)R_{n,k}||&=\frac{1}{n}\sqrt{n}\sum_{k=1}^n||A(\beta, Z_k)R_{n,k}||\\
        &\leq \sqrt{n}\max_{k\leq n}|R_{n,k}|\frac{1}{n}\sum_{k=1}^n||A(\beta, Z_k)||\\
        &=o_p(1)O_p(1).
    \end{split}
\end{equation*}
Because $\hat{\theta}^{\ast}$ is bounded, it is sufficient to consider $||A(\beta, Z_k)\mu(\beta, Z_k))||$, for which we have by Markov's inequality
\begin{equation*}
    \begin{split}
        &\sum_{k=1}^n P(||\frac{1}{\sqrt{n}}A(\beta, Z_k)(\mu(\beta, Z_k))||>\varepsilon)\\
        =&n P(||A(\beta, Z_1)(\mu(\beta, Z_1))||>\varepsilon\sqrt{n})\\
        \leq& n\frac{E(||A(\beta, Z_1)(\mu(\beta, Z_1))||^{2+\delta})}{\varepsilon^{2+\delta}n^{1+\frac{\delta}{2}}}\\
        =&\frac{E(||A(\beta, Z_1)(\mu(\beta, Z_1))||^{2+\delta})}{\varepsilon^{2+\delta}n^{\frac{\delta}{2}}} \to 0
    \end{split}
\end{equation*}
as $n\to \infty,$ due to the assumption that $E(||A(\beta, Z_1)(\mu(\beta, Z_1))||^{2+\delta})<\infty.$

To prove (b), rewrite $\frac{1}{\sqrt{n}} U_n^B(\beta_0)$ as
\[\frac{1}{\sqrt{n}}\sum_{k=1}^n A(\beta_0, Z_k)(\hat{\theta}_{n,k} -\mu(\beta_0, Z_k)) +\frac{1}{\sqrt{n}}\sum_{k=1}^n (M_{nk}-1) A(\beta_0, Z_k)(\hat{\theta}_{n,k} -\mu(\beta_0, Z_k))\]
which is a continuous function of the vector
\[L_n=\frac{1}{\sqrt{n}}\left(\sum_{k=1}^n A(\beta_0, Z_k)(\hat{\theta}_{n,k} -\mu(\beta_0, Z_k)), \sum_{k=1}^n (M_{nk}-1) A(\beta_0, Z_k)(\hat{\theta}_{n,k} -\mu(\beta_0, Z_k))\right).\]
The first component converges in distribution to $Q\sim N_q(0, \Sigma)$, and by (a) the second converges in distribution to $Q'\sim N_q(0, \Tilde{\Sigma})$ in probability given the data. By Lemma 1 in \citet{munko2024conditional} this implies the unconditional convergence $L_n\xrightarrow{d}(Q,Q')$ for independent $Q$ and $Q'$ and therefore $\frac{1}{\sqrt{n}} U_n^B(\beta_0)\xrightarrow{d}Q+Q'.$
\end{proof}

\begin{proof}[Proof of Theorem \ref{weakconvbootstrapestimator}]
    To establish existence and consistency of the bootstrap estimator, we apply Theorem 2.5 of \citet{jacod2018review}. Because the arguments are identical to the proof of Theorem \ref{overgaard main variant}, we omit some of the explicit calculations. We can again assume w.l.o.g. that the dominated integrability conditions hold for $G=\overline{B}_{\varepsilon}(\beta_0).$ It follows from Equation \eqref{kernelderivativematrix}, that the kernel fulfills Condition \ref{uniformmomentcond} (2) due to the dominated integrability conditions and because the influence functions $\dot\phi(.)$ and $\ddot\phi(.,.)$ are bounded. Because the Frobenius norm is induced by an inner product, Condition \ref{uniformmomentcond} (3) is fulfilled as well. Consider the resampling counterpart of \eqref{derivativeUstatistic}
    \begin{equation*}
            \begin{split}
        \frac{\partial}{\partial \beta}(\frac{1}{n}U_n^{B\ast}(\beta))= & \frac{1}{n}\sum_{k=1}^n M_{nk}\frac{\partial}{\partial\beta}(A(\beta, Z_k))(\hat{\theta}^{\ast}_{n,k} -\mu(\beta, Z_k)) \\
        & - \frac{1}{n} \sum_{k=1}^n M_{nk}A(\beta, Z_k)\frac{\partial}{\partial\beta}\mu(\beta, Z_k).
    \end{split}
    \end{equation*}
    By Theorem \ref{Uniform LLN U and V} (c)
    \begin{equation*}
        \sup_{\beta\in G}||\frac{\partial}{\partial \beta}(\frac{1}{n}U_n^{B\ast}(\beta))-M(\beta_0)||\xrightarrow{p}0
    \end{equation*}
    as $n\to\infty.$ For the resampling matrix
    \begin{equation*}
        \begin{split}
        \frac{\partial}{\partial \beta}(\frac{1}{n}U_n^B(\beta))= & \frac{1}{n}\sum_{k=1}^n \frac{\partial}{\partial\beta}M_{nk}(A(\beta, Z_k))(\hat{\theta}_{n,k} -\mu(\beta, Z_k)) \\
        & - \frac{1}{n} \sum_{k=1}^n M_{nk}A(\beta, Z_k)\frac{\partial}{\partial\beta}\mu(\beta, Z_k)
    \end{split}
    \end{equation*}
    we have
    \begin{equation*}
        \begin{split}
            &\sup_{\beta\in G}||\frac{\partial}{\partial \beta}(\frac{1}{n}U_n^{B\ast}(\beta))-\frac{\partial}{\partial \beta}(\frac{1}{n}U_n^B(\beta))||\\
            =&\sup_{\beta\in G}||\frac{1}{n}\sum_{k=1}^nM_{nk}\frac{\partial}{\partial\beta}(A(\beta, Z_k))(\hat{\theta}^{\ast}_{n,k} -\hat{\theta}_{n,k})||\\
            \leq& \max_{k\leq n}|R_{n,k}|\frac{1}{n}\sum_{k=1}^n\sup_{\beta\in G}||M_{nk}\frac{\partial}{\partial\beta}A(\beta, Z_k)||=o_p(1)O_p(1)
        \end{split}
    \end{equation*}
    by the bootstrap law of large numbers, which is enclosed in the bootstrap law of large numbers for U-Statistics (Theorem 13.3.1 in \citet{borovskikh1996ustatistics}). This establishes existence and weak consistency of $\Hat{\beta}_n^B.$
    
    Furthermore, Theorem 2.11 of \citet{jacod2018review} together with Lemma \ref{weak covergence Boot equation} (b) imply the unconditional weak convergence
    \begin{equation*}
        \sqrt{n}(\hat{\beta}_n^B-\beta_0) \xrightarrow{d} N_q(0,M^{-1}(\Sigma+\Tilde{\Sigma})(M^{-1})^T).
    \end{equation*}
    Additionally, the following asymptotic representations hold
    \begin{equation}\label{asy1}
        \sqrt{n}(\hat{\beta}_n-\beta_0)= -(\frac{\partial}{\partial \beta}(\frac{1}{n}U_n(\beta)))^{-1}\frac{1}{\sqrt{n}}U_n(\beta_0)+o_p(1),
    \end{equation}
        \begin{equation}\label{asy2}
        \sqrt{n}(\hat{\beta}_n^B-\beta_0)= -(\frac{\partial}{\partial \beta}(\frac{1}{n}U_n^B(\beta)))^{-1}\frac{1}{\sqrt{n}}U_n^B(\beta_0)+o_p(1).
    \end{equation}
    For a more compact notation we introduce the abbreviations $W_n:=\frac{\partial}{\partial \beta}(\frac{1}{n}U_n(\beta))$ and $W_n^B:=\frac{\partial}{\partial \beta}(\frac{1}{n}U_n^B(\beta)).$ Subtracting \eqref{asy1} from \eqref{asy2} we get
    \begin{equation*}
        \begin{split}
            \sqrt{n}(\hat{\beta}_n^B-\hat{\beta}_n)=&-(W_n^B)^{-1}\frac{1}{\sqrt{n}}U_n^B(\beta_0)\\
            &-W_n^{-1}\frac{1}{\sqrt{n}}U_n(\beta_0)+o_p(1)\\
            =&-(W_n^B)^{-1}\Big(\frac{1}{\sqrt{n}}\sum_{k=1}^n (M_{nk}-1) A(\beta_0, Z_k)(\hat{\theta}_{n,k} -\mu(\beta_0, Z_k))\\
            &+\frac{1}{\sqrt{n}}U_n(\beta_0) +W_n^B W_n^{-1}\frac{1}{\sqrt{n}}U_n(\beta_0)\Big)+o_p(1)\\
            =&-(W_n^B)^{-1}\frac{1}{\sqrt{n}}\sum_{k=1}^n (M_{nk}-1) A(\beta_0, Z_k)(\hat{\theta}_{n,k} -\mu(\beta_0, Z_k))\\
            &-(W_n^B)^{-1}(I-W_n^B W_n^{-1})\frac{1}{\sqrt{n}}U_n(\beta_0)+o_p(1).
        \end{split}
    \end{equation*}
    By Theorem \ref{weak covergence Boot equation} (a) and Slutzkys lemma
    \begin{equation*}
        -(W_n^B)^{-1}\frac{1}{\sqrt{n}}\sum_{k=1}^n (M_{nk}-1) A(\beta_0, Z_k)(\hat{\theta}_{n,k} -\mu(\beta_0, Z_k))\xrightarrow{d}N_q(0,M^{-1}\Tilde{\Sigma}(M^{-1})^T)
    \end{equation*}
    given the data in probability. Because $W_n$ and $W_n^B$ converge in probability to the same limit, $(I-W_n^B W_n^{-1})$ converges in probability to 0. This implies
    \begin{equation*}
        \sqrt{n}(\hat{\beta}_n^B-\hat{\beta}_n)\xrightarrow{d} N_q(0,M^{-1}\Tilde{\Sigma}(M^{-1})^T)
    \end{equation*}
given the data in probability.
\end{proof}

\begin{proof}[Proof of Theorem \ref{bootstraptest}]
    Due to Theorem \ref{weakconvbootstrapestimator}, it suffices to show that  $\Hat{M}^{B}_n$ and $\Hat{\Sigma}^{\text{HW, B}}_n$ converge in (conditional) probability to the same limit as their non-bootstrap versions. The proof of Lemma \ref{HW convergence} is also valid for $\Hat{\Sigma}^{\text{HW, B}}_n$, if some minor adjustments are made. First notice, that because $\hat{\beta}_n^B$ is consistent for $\beta_0$, the probability of the event $\{\Hat{\beta}^B_n \in \overline{B}_{\varepsilon}(\beta_0)\}$ tends to 1 as well. We also use the bootstrap uniform law of large numbers (Theorem \ref{Uniform LLN U and V} (c)) for U-Statistics instead of the usual uniform law of large numbers for U-Statistics, which is possible because $A(\beta,Z)$ is locally dominated square integrable. Finally, we have for the bootstrap remainder $\max_{k\leq n}|R_{n,k}^B|\leq\max_{k\leq n}|R_{n,k}|,$ because we resample from the already calculated pseudo-observations. Similarly, Lemma \ref{M estimator} is valid for $\hat{M}_n^{B}$ under the same assumptions, because the uniform law of large for bootstrap means does not require the existence of moments higher than the first.
\end{proof}

\newpage
\section{Additional simulation results}\label{appendix:C add simulation}
\subsection{Further simulation results under the alternative hypothesis}

The description of the simulation setup can be found in Section~\ref{sec:simulation and data example}. The following results complement the results shown in the main manuscript.

\begin{table}[ht!]
\centering
\caption{Power of the tests under $H_1^{(2)}:C^{(2)}\beta_0\ne 0$, the true effect is $\delta_2=1.$}
\vspace{2mm}
\label{table:power like veteran celltype+1}
\begin{tabular}{|rr|r|rrrrr|}
\hline
\multicolumn{1}{|l}{} & \multicolumn{1}{l}{} & \multicolumn{1}{|l|}{}   & \multicolumn{5}{c|}{$H_0^{(2)}:C^{(2)}\beta_0=0$}                       \\
$n$                   & $\vartheta$                & $\delta_1$            & $\varphi^{\text{Corr}}$ & $\varphi^{\text{HW}}$ & $\varphi^{B_{\text{HW}}}$ & $\varphi^{\text{HC3}}$ & $\varphi^{B_{\text{HC3}}}$  \\ \hline
80 & $\infty$ & 0.0 & 20.4 & 20.4 & 0.6 & 11.8 & 0.4 \\ 
  137 & $\infty$ & 0.0 & 40.1 & 40.1 & 31.5 & 33.2 & 30.8 \\ 
  200 & $\infty$ & 0.0 & 59.4 & 59.4 & 57.6 & 55.3 & 57.2 \\ 
  80 & 730 & 0.0 & 16.4 & 15.6 & 15.9 & 8.0 & 16.5 \\ 
  137 & 730 & 0.0 & 36.8 & 36.3 & 34.0 & 28.8 & 34.1 \\ 
  200 & 730 & 0.0 & 57.1 & 56.7 & 55.7 & 51.4 & 55.7 \\ 
  80 & 365 & 0.0 & 14.8 & 13.0 & 17.2 & 5.1 & 16.4 \\ 
  137 & 365 & 0.0 & 32.3 & 31.2 & 32.0 & 24.4 & 32.3 \\ 
  200 & 365 & 0.0 & 52.8 & 52.2 & 52.1 & 47.2 & 52.6 \\ 
  80 & $\infty$ & 0.5 & 17.0 & 17.0 & 0.1 & 9.1 & 0.2 \\ 
  137 & $\infty$ & 0.5 & 35.4 & 35.4 & 18.6 & 26.2 & 18.6 \\ 
  200 & $\infty$ & 0.5 & 53.3 & 53.3 & 48.4 & 47.8 & 49.0 \\ 
  80 & 730 & 0.5 & 13.8 & 13.2 & 13.2 & 5.7 & 12.8 \\ 
  137 & 730 & 0.5 & 31.1 & 30.6 & 26.8 & 22.8 & 26.9 \\ 
  200 & 730 & 0.5 & 49.8 & 49.5 & 46.5 & 44.0 & 47.4 \\ 
  80 & 365 & 0.5 & 10.8 & 9.5 & 12.3 & 4.1 & 13.0 \\ 
  137 & 365 & 0.5 & 28.2 & 27.2 & 26.8 & 19.8 & 26.4 \\ 
  200 & 365 & 0.5 & 45.1 & 44.4 & 43.5 & 39.2 & 43.6 \\ 
  80 & $\infty$ & 1.0 & 15.8 & 15.9 & 0.2 & 9.3 & 0.2 \\ 
  137 & $\infty$ & 1.0 & 27.0 & 27.0 & 6.9 & 19.3 & 7.6 \\ 
  200 & $\infty$ & 1.0 & 45.2 & 45.2 & 34.4 & 37.5 & 33.8 \\ 
  80 & 730 & 1.0 & 11.2 & 10.4 & 9.9 & 4.2 & 9.3 \\ 
  137 & 730 & 1.0 & 24.5 & 24.2 & 19.5 & 17.1 & 19.5 \\ 
  200 & 730 & 1.0 & 41.5 & 41.3 & 37.1 & 33.6 & 35.7 \\ 
  80 & 365 & 1.0 & 9.1 & 8.2 & 10.8 & 2.5 & 10.0 \\ 
  137 & 365 & 1.0 & 21.9 & 21.0 & 18.6 & 13.8 & 18.8 \\ 
  200 & 365 & 1.0 & 35.7 & 35.0 & 32.5 & 29.7 & 33.2 \\ 
   \hline
\end{tabular}
\end{table}

\begin{table}[ht!]
\centering
\caption{Power of the tests under $H_1^{(1)}:C^{(1)}\beta_0\ne 0$, the true effect is $\delta_1=0.5.$}
\vspace{2mm}
\label{table:power like veteran trt 0.5}
\begin{tabular}{|rr|r|rrrrr|}
\hline
\multicolumn{1}{|l}{} & \multicolumn{1}{l}{} & \multicolumn{1}{|l|}{}   & \multicolumn{5}{c|}{$H_0^{(2)}:C^{(2)}\beta_0=0$}                       \\
$n$                   & $\vartheta$                & $\delta_1$            & $\varphi^{\text{Corr}}$ & $\varphi^{\text{HW}}$ & $\varphi^{B_{\text{HW}}}$ & $\varphi^{\text{HC3}}$ & $\varphi^{B_{\text{HC3}}}$  \\ \hline
80 & $\infty$ & 0.0 & 20.9 & 20.9 & 4.4 & 13.9 & 3.9 \\ 
  137 & $\infty$ & 0.0 & 27.5 & 27.5 & 21.8 & 26.2 & 23.8 \\ 
  200 & $\infty$ & 0.0 & 39.4 & 39.4 & 35.3 & 36.5 & 35.7 \\ 
  80 & 730 & 0.0 & 17.5 & 17.3 & 13.9 & 12.8 & 13.8 \\ 
  137 & 730 & 0.0 & 27.6 & 27.6 & 22.8 & 24.0 & 22.1 \\ 
  200 & 730 & 0.0 & 37.0 & 36.9 & 33.3 & 35.0 & 33.7 \\ 
  80 & 365 & 0.0 & 17.0 & 16.4 & 15.4 & 11.5 & 14.7 \\ 
  137 & 365 & 0.0 & 26.4 & 26.2 & 21.9 & 22.4 & 21.2 \\ 
  200 & 365 & 0.0 & 35.1 & 34.9 & 31.4 & 33.2 & 32.3 \\ 
  80 & $\infty$ & -1.0 & 19.3 & 19.3 & 4.3 & 14.7 & 4.4 \\ 
  137 & $\infty$ & -1.0 & 28.6 & 28.6 & 22.4 & 26.2 & 23.6 \\ 
  200 & $\infty$ & -1.0 & 38.0 & 38.0 & 34.4 & 37.3 & 35.9 \\ 
  80 & 730 & -1.0 & 18.1 & 17.8 & 14.6 & 13.2 & 14.9 \\ 
  137 & 730 & -1.0 & 26.9 & 26.8 & 22.1 & 25.0 & 22.9 \\ 
  200 & 730 & -1.0 & 36.9 & 36.8 & 33.3 & 34.7 & 33.5 \\ 
  80 & 365 & -1.0 & 17.1 & 16.6 & 16.3 & 11.6 & 15.8 \\ 
  137 & 365 & -1.0 & 26.0 & 25.7 & 21.6 & 21.9 & 21.3 \\ 
  200 & 365 & -1.0 & 34.3 & 34.1 & 30.9 & 31.9 & 30.8 \\ 
  80 & $\infty$ & 1.0 & 17.2 & 17.2 & 2.3 & 13.5 & 2.4 \\ 
  137 & $\infty$ & 1.0 & 27.8 & 27.8 & 21.0 & 23.4 & 19.5 \\ 
  200 & $\infty$ & 1.0 & 37.0 & 37.0 & 32.6 & 33.9 & 32.0 \\ 
  80 & 730 & 1.0 & 15.9 & 15.7 & 13.1 & 11.1 & 12.4 \\ 
  137 & 730 & 1.0 & 25.7 & 25.6 & 20.2 & 22.0 & 20.1 \\ 
  200 & 730 & 1.0 & 33.7 & 33.6 & 29.6 & 31.0 & 29.2 \\ 
  80 & 365 & 1.0 & 15.6 & 15.2 & 15.0 & 9.6 & 13.4 \\ 
  137 & 365 & 1.0 & 23.2 & 22.9 & 18.9 & 20.3 & 19.1 \\ 
  200 & 365 & 1.0 & 31.2 & 31.1 & 27.2 & 29.0 & 27.2 \\ 
   \hline
\end{tabular}
\end{table}

\begin{table}[ht!]
\centering
\caption{Power of the tests under $H_1^{(1)}:C^{(1)}\beta_0\ne 0$, the true effect is $\delta_1=1.$}
\vspace{2mm}
\label{table:power like veteran trt 1}
\begin{tabular}{|rr|r|rrrrr|}
\hline
\multicolumn{1}{|l}{} & \multicolumn{1}{l}{} & \multicolumn{1}{|l|}{}   & \multicolumn{5}{c|}{$H_0^{(2)}:C^{(2)}\beta_0=0$}                       \\
$n$                   & $\vartheta$                & $\delta_1$            & $\varphi^{\text{Corr}}$ & $\varphi^{\text{HW}}$ & $\varphi^{B_{\text{HW}}}$ & $\varphi^{\text{HC3}}$ & $\varphi^{B_{\text{HC3}}}$  \\ \hline
80 & $\infty$ & 0.0 & 51.1 & 51.1 & 9.2 & 43.3 & 8.6 \\ 
  137 & $\infty$ & 0.0 & 76.3 & 76.3 & 68.4 & 72.6 & 68.1 \\ 
  200 & $\infty$ & 0.0 & 90.4 & 90.4 & 88.6 & 87.9 & 86.9 \\ 
  80 & 730 & 0.0 & 48.2 & 47.9 & 38.4 & 38.7 & 38.2 \\ 
  137 & 730 & 0.0 & 72.2 & 72.0 & 65.1 & 68.4 & 65.5 \\ 
  200 & 730 & 0.0 & 87.6 & 87.6 & 85.1 & 86.7 & 85.5 \\ 
  80 & 365 & 0.0 & 44.9 & 44.2 & 38.7 & 35.3 & 38.2 \\ 
  137 & 365 & 0.0 & 71.3 & 70.9 & 64.8 & 66.0 & 63.7 \\ 
  200 & 365 & 0.0 & 86.5 & 86.4 & 83.0 & 83.7 & 82.3 \\ 
  80 & $\infty$ & -1.0 & 52.6 & 52.6 & 10.9 & 44.2 & 11.5 \\ 
  137 & $\infty$ & -1.0 & 75.9 & 75.9 & 69.0 & 73.0 & 69.3 \\ 
  200 & $\infty$ & -1.0 & 89.9 & 89.9 & 87.9 & 89.2 & 87.9 \\ 
  80 & 730 & -1.0 & 49.0 & 48.7 & 43.7 & 39.8 & 42.1 \\ 
  137 & 730 & -1.0 & 73.3 & 73.1 & 67.0 & 69.9 & 67.1 \\ 
  200 & 730 & -1.0 & 88.1 & 88.1 & 85.0 & 87.1 & 85.9 \\ 
  80 & 365 & -1.0 & 45.7 & 45.0 & 43.2 & 35.5 & 41.5 \\ 
  137 & 365 & -1.0 & 71.2 & 70.9 & 65.3 & 66.2 & 65.3 \\ 
  200 & 365 & -1.0 & 84.4 & 84.2 & 81.3 & 84.0 & 83.2 \\ 
  80 & $\infty$ & 1.0 & 46.2 & 46.2 & 4.0 & 38.2 & 4.2 \\ 
  137 & $\infty$ & 1.0 & 71.3 & 71.3 & 58.0 & 65.4 & 56.8 \\ 
  200 & $\infty$ & 1.0 & 85.0 & 85.0 & 81.5 & 84.3 & 82.4 \\ 
  80 & 730 & 1.0 & 42.1 & 42.0 & 33.3 & 31.1 & 30.9 \\ 
  137 & 730 & 1.0 & 68.5 & 68.4 & 60.1 & 61.6 & 58.1 \\ 
  200 & 730 & 1.0 & 83.7 & 83.6 & 79.4 & 81.2 & 79.3 \\ 
  80 & 365 & 1.0 & 37.3 & 36.9 & 31.8 & 26.5 & 29.8 \\ 
  137 & 365 & 1.0 & 63.4 & 63.0 & 55.7 & 59.3 & 56.4 \\ 
  200 & 365 & 1.0 & 80.3 & 80.2 & 75.6 & 78.1 & 76.0 \\ 
   \hline
\end{tabular}
\end{table}

\newpage
\phantom{X}
\newpage
\phantom{X}
\newpage

\subsection{Second simulation study: Including interaction effects in the linear predictor when modeling the survival function at a fixed time point}\label{sec:simu with interaction}
The setup is similar to the simulation study of the main paper, but we included an interaction effect.
As estimand of interest, we chose the conditional survival function modeled as 
$S(t_0|Z)=\mu(\beta^T_0 Z)$ 
with the inverse logit $g(x)=\frac{\exp(x)}{1+\exp(x)}$ as a response function, $\beta_0 \in \mathbb{R}^5$ and $Z^T = (1,Z_1, \dots, Z_4)$.
The covariates $Z_1$ and $Z_2$ are independently $Bin(1, 0.5)$-distributed, $Z_3 := Z_1 Z_2$ models an interaction term, and $Z_4\sim U(0,1)$ models the influence of a continuous covariate. Survival times were generated from a Weibull distribution with density 
\[f(t)=\frac{a}{\lambda}\left(\frac{t}{a}\right)^{a-1}e^{-\left(\frac{t}{\lambda}\right)^a}\mathbbm{1}\{t>0\},\]
fixed shape parameter $a=2$, and individual scale parameters 
\[\lambda=t_0(-\log(\mu(\beta_0^TZ))^{-\frac{1}{a}},\]
which are obtained by solving the survival function $\mu(\beta_0^TZ)=\exp(-(\frac{t_0}{\lambda})^{a})$ for $\lambda.$ We considered sample sizes $n\in \{80,140,200\}.$ The time point of interest was chosen as $t_0=1.$
As for censoring, we simulated $C_i \sim U(0,3), U(0,5)$ as well as settings with no censoring, $C_i = \infty$. We summarize this censoring scheme in the vector $\vartheta= (3,5,\infty).$ We used the function $A(\beta, Z)=\frac{\partial}{\partial\beta}\mu(\beta^TZ).$ The parameter vector was set to $\beta_0=(0.3,0,0,\delta,0.1)$  with $\delta\in \{0,1,2\}.$ We considered the hypothesis matrices
\[C^{(1)}=\begin{pmatrix}
0 & 1 & 0 & 0 & 0 \\
\end{pmatrix},\]
\[C^{(2)}=\begin{pmatrix}
0 & 1 & 0 & 0 & 0 \\
0 & 0 & 0 & 1 & 0 
\end{pmatrix}, \]
and
\[C^{(3)}=\begin{pmatrix}
0 & 1 & 0 & 0 & 0 \\
0 & 0 & 1 & 0 & 0 \\
0 & 0 & 0 & 1 & 0 
\end{pmatrix} .\]
For $\delta=0$ all null hypotheses are true, otherwise only $H_{0}^{(1)}:C^{(1)}\beta_0=0$ holds. All hypotheses were tested at the significance level of $\alpha=0.05$ using the same tests as in Section~5.
The simulations were conducted in the R computing environment, version 4.4.0 \citep{citeR}, each with $n_{\text{sim}}=5000$ simulation runs. The random quantiles $q^{B}_{n,1-\alpha}$ were determined via $B=1000$ bootstrap runs per simulation step. 

\begin{sidewaystable}
\centering
\caption{Rejection rates for the hypothesis matrices  $C^{(1)}$, $C^{(2)}$ and $C^{(3)}$.}
\vspace{2mm}
\label{table:results additional simu setup}
\scriptsize
\begin{threeparttable}
\begin{tabular}{|rr|r|rrrrr|rrrrr|rrrrr|}
  \hline
  \multicolumn{1}{|l}{} & \multicolumn{1}{l}{} & \multicolumn{1}{|l|}{}   & \multicolumn{5}{c|}{$H_0^{(1)}:C^{(1)}\beta_0=0$}  & \multicolumn{5}{c|}{$H_0^{(2)}:C^{(2)}\beta_0=0$} & \multicolumn{5}{c|}{$H_0^{(3)}:C^{(3)}\beta_0=0$}                     \\
$n$ & $\vartheta$ & $\delta$ & $\varphi^{\text{Corr}}$ & $\varphi^{\text{HW}}$ & $\varphi^{B_{\text{HW}}}$ & $\varphi^{\text{HC3}}$ & $\varphi^{B_{\text{HC3}}}$ & $\varphi^{\text{Corr}}$ & $\varphi^{\text{HW}}$ & $\varphi^{B_{\text{HW}}}$ & $\varphi^{\text{HC3}}$ & $\varphi^{B_{\text{HC3}}}$ & $\varphi^{\text{Corr}}$ & $\varphi^{\text{HW}}$ & $\varphi^{B_{\text{HW}}}$ & $\varphi^{\text{HC3}}$ & $\varphi^{B_{\text{HC3}}}$ \\ 
  \hline
80 & $\infty$ & 0.0 & 5.5 & 5.5 & 2.9 & 3.4 & 2.4 & 5.2 & 5.2 & 1.7 & 2.9 & 1.2 & 3.8 & 3.8 & 1.1 & 1.9 & 1.0 \\ 
  140 & $\infty$ & 0.0 & 4.6 & 4.6 & 4.4 & 3.6 & 4.2 & 4.8 & 4.8 & 4.6 & 3.7 & 4.3 & 4.5 & 4.5 & 4.7 & 3.2 & 4.5 \\ 
  200 & $\infty$ & 0.0 & 4.9 & 4.9 & 4.8 & 5.1 & 6.0 & 4.9 & 4.9 & 5.1 & 4.1 & 5.2 & 4.6 & 4.6 & 5.0 & 3.9 & 5.3 \\ 
  80 & 5 & 0.0 & 4.6 & 4.3 & 6.2 & 3.1 & 6.7 & 3.7 & 3.4 & 6.1 & 2.1 & 6.1 & 3.1 & 2.6 & 5.0 & 1.5 & 5.4 \\ 
  140 & 5 & 0.0 & 5.4 & 5.2 & 5.7 & 3.8 & 5.4 & 4.6 & 4.4 & 5.2 & 3.4 & 5.6 & 4.0 & 3.8 & 5.3 & 2.4 & 4.9 \\ 
  200 & 5 & 0.0 & 5.1 & 5.0 & 5.2 & 4.9 & 5.7 & 5.1 & 4.9 & 5.3 & 4.4 & 5.6 & 4.7 & 4.4 & 5.4 & 3.8 & 5.3 \\ 
  80 & 3 & 0.0 & 4.5 & 4.0 & 7.4 & 2.3 & 7.2 & 3.8 & 2.9 & 7.0 & 1.5 & 6.9 & 2.4 & 1.7 & 5.6 & 0.7 & 5.8 \\ 
  140 & 3 & 0.0 & 5.3 & 5.1 & 6.3 & 3.6 & 5.7 & 4.8 & 4.4 & 6.3 & 3.3 & 5.7 & 4.0 & 3.4 & 6.0 & 2.3 & 5.8 \\ 
  200 & 3 & 0.0 & 4.8 & 4.7 & 5.5 & 4.7 & 5.9 & 5.0 & 4.8 & 5.8 & 4.4 & 6.1 & 4.7 & 4.2 & 5.9 & 3.5 & 6.2 \\ 
  \hline
  80 & $\infty$ & 1.0 & 5.2 & 5.2 & 2.5 & 3.7 & 2.3 & 19.7 & 19.7 & 3.4 & 14.0 & 2.7 & 18.6 & 18.6 & 3.1 & 12.7 & 2.4 \\ 
  140 & $\infty$ & 1.0 & 4.5 & 4.5 & 4.3 & 4.2 & 4.8 & 34.0 & 34.0 & 28.0 & 30.1 & 27.8 & 38.8 & 38.8 & 33.2 & 33.9 & 33.1 \\ 
  200 & $\infty$ & 1.0 & 4.7 & 4.7 & 4.6 & 5.1 & 5.6 & 49.6 & 49.6 & 48.8 & 47.3 & 49.1 & 57.2 & 57.2 & 57.6 & 55.1 & 58.9 \\ 
  80 & 5 & 1.0 & 4.5 & 4.3 & 6.9 & 3.0 & 6.6 & 13.3 & 12.2 & 18.0 & 7.8 & 12.6 & 12.3 & 10.5 & 18.7 & 5.8 & 13.2 \\ 
  140 & 5 & 1.0 & 5.1 & 4.9 & 5.3 & 3.9 & 5.1 & 28.3 & 27.6 & 31.5 & 23.8 & 29.8 & 31.3 & 30.0 & 35.5 & 24.4 & 34.2 \\ 
  200 & 5 & 1.0 & 5.2 & 5.2 & 5.2 & 4.6 & 5.7 & 41.9 & 41.2 & 43.8 & 39.6 & 45.1 & 48.3 & 47.5 & 51.7 & 45.6 & 52.8 \\ 
  80 & 3 & 1.0 & 4.1 & 3.8 & 7.1 & 2.3 & 7.3 & 10.7 & 8.7 & 16.8 & 4.7 & 11.5 & 9.1 & 6.4 & 15.9 & 3.1 & 11.2 \\ 
  140 & 3 & 1.0 & 4.9 & 4.7 & 6.0 & 3.9 & 5.8 & 23.7 & 22.3 & 28.4 & 18.6 & 26.0 & 24.7 & 22.1 & 31.2 & 18.1 & 28.9 \\ 
  200 & 3 & 1.0 & 4.6 & 4.4 & 5.0 & 4.8 & 5.8 & 37.2 & 36.1 & 40.1 & 34.2 & 40.4 & 42.0 & 40.1 & 47.0 & 37.9 & 47.0 \\
  \hline
  80 & $\infty$ & 2.0 & 4.9 & 4.9 & 2.1 & 3.1 & 1.8 & 51.3 & 51.3 & 10.6 & 43.5 & 7.5 & 51.3 & 51.3 & 11.9 & 41.5 & 6.3 \\ 
  140 & $\infty$ & 2.0 & 4.9 & 4.9 & 4.6 & 4.0 & 4.4 & 81.2 & 81.2 & 31.5 & 74.8 & 28.6 & 87.8 & 87.8 & 36.0 & 81.6 & 33.0 \\ 
  200 & $\infty$ & 2.0 & 4.8 & 4.8 & 4.9 & 4.9 & 5.4 & 94.1 & 94.1 & 67.0 & 92.8 & 65.2 & 98.2 & 98.2 & 71.5 & 97.6 & 70.5 \\ 
  80 & 5 & 2.0 & 4.7 & 4.5 & 8.0 & 2.3 & 6.2 & 25.5 & 22.1 & 33.1 & 14.8 & 15.1 & 25.5 & 21.2 & 33.7 & 12.5 & 15.9 \\ 
  140 & 5 & 2.0 & 4.7 & 4.6 & 5.9 & 3.7 & 5.6 & 59.8 & 57.9 & 64.2 & 50.8 & 51.2 & 63.7 & 61.0 & 68.4 & 53.3 & 56.1 \\ 
  200 & 5 & 2.0 & 4.8 & 4.7 & 5.5 & 3.6 & 4.5 & 83.6 & 82.9 & 85.4 & 79.2 & 80.6 & 88.9 & 88.1 & 90.0 & 85.3 & 85.7 \\ 
  80 & 3 & 2.0 & 3.6 & 3.3 & 8.7 & 2.1 & 7.3 & 18.3 & 14.5 & 27.6 & 9.7 & 12.9 & 17.3 & 12.5 & 27.1 & 7.0 & 13.1 \\ 
  140 & 3 & 2.0 & 4.7 & 4.4 & 6.7 & 3.8 & 5.9 & 45.6 & 42.2 & 51.5 & 37.1 & 37.7 & 47.6 & 42.6 & 54.2 & 36.8 & 41.4 \\ 
  200 & 3 & 2.0 & 5.0 & 4.8 & 6.0 & 4.2 & 5.5 & 74.1 & 72.0 & 76.6 & 67.5 & 67.1 & 78.7 & 76.0 & 80.7 & 73.0 & 72.8 \\  
   \hline
\end{tabular}
\begin{tablenotes}
    \item The $2\times 2$ block in the lower right of the table contains the settings where the alternative hypotheses are true, the remaining entries contain results where the null hypotheses are true.
\end{tablenotes}
\end{threeparttable}

\end{sidewaystable}
Table \ref{table:results additional simu setup} displays the results for the additional simulations with an included interaction effect. The behavior of the tests is very similar to the behavior observed in the simulation study of the main paper. The bootstrap-based tests however are rather liberal in many scenarios, whereas the test based on the corrected variance estimator controls the type-I error rate well in most scenarios, even in small samples. We therefore recommend its use over the tests based on the Huber-White and HC3 estimators as well as the naive bootstrap tests in models with interaction effects.

\end{document}